\documentclass{aa}  

\usepackage{graphicx}
\usepackage{txfonts}
\usepackage{color}
\usepackage[colorlinks=true, allcolors=blue]{hyperref}
\usepackage{multirow}
\usepackage{booktabs}

\usepackage{natbib}
\bibpunct{(}{)}{;}{a}{}{,}

\newcommand{\uergcm}[1]{erg cm$^{-2}$ s$^{-1}$}
\newcommand{\ergcm}[1]{$\times 10^{#1}$ erg cm$^{-2}$ s$^{-1}$}
\newcommand{\oergcm}[1]{$10^{#1}$ erg cm$^{-2}$ s$^{-1}$}
\newcommand{\ergs}[1]{$\times 10^{#1}$ erg s$^{-1}$}
\newcommand{\oergs}[1]{$10^{#1}$ erg s$^{-1}$}

\newcommand{\ohcm}[1]{\ensuremath{10^{#1}$ cm$^{-2}}}

\newcommand{\oexpo}[1]{$10^{#1}$}
\newcommand{\kms}{km s$^{-1}$\xspace}
\newcommand{\nh}{\ensuremath{N_\mathrm{H}}} 
\newcommand{\lx}{\ensuremath{L_\mathrm{X}}}

\newcommand{\fx}{\hbox{F$_{\rm x}$}}
\newcommand{\cts}{cts s$^{-1}$}

\newcommand{\Halpha}{H${\alpha}$\xspace}
\newcommand{\ltsima}{$\buildrel < \over \sim$}
\newcommand{\lsim}{\lower.5ex\hbox{\ltsima}}
\newcommand{\gtsima}{$\buildrel > \over \sim$}
\newcommand{\gsim}{\lower.5ex\hbox{\gtsima}}
\newcommand{\msun}{M$_{\odot}$\xspace}

\newcommand{\xspec}{\texttt{XSPEC}\xspace}
\newcommand{\eSASS}{\texttt{eSASS}\xspace}

\newcommand{\ROSAT}{{\it ROSAT}\xspace}
\newcommand{\gaia}{{\it Gaia}\xspace}
\newcommand{\swift}{{\it Swift}\xspace}
\newcommand{\xmm}{{\it XMM-Newton}\xspace}
\newcommand{\nus}{{\it NuSTAR}\xspace}

\newcommand{\ero}{\mbox{eROSITA}\xspace}
\newcommand{\srg}{{\it SRG}\xspace}

\newcommand{\src}{\mbox{eRASSU\,J052914.9$-$662446}\xspace}

\begin{document} 

\title{Broadband study and the discovery of pulsations from the Be/X-ray binary \src in the Large Magellanic Cloud}

\author{C. Maitra\inst{\ref{mpe}} \and
        D. Kaltenbrunner\inst{\ref{mpe}} \and
        F. Haberl\inst{\ref{mpe}} \and
        D.\,A.\,H. Buckley\inst{\ref{salt},\ref{uct}} \and
        I.\,M. Monageng\inst{\ref{salt},\ref{uct}} \and
        A.\,Udalski\inst{\ref{uw}} \and
        S.\,Carpano\inst{\ref{mpe}} \and
        J.\,B.\,Coley\inst{\ref{howard},\ref{gsfc}} \and
        V.\,Doroshenko\inst{\ref{iaat}}  \and
        L. Ducci\inst{\ref{iaat}}  \and
        C. Malacaria\inst{\ref{usra}}  \and
        O. K\"onig\inst{\ref{uerl}} \and
        A. Santangelo\inst{\ref{iaat}}  \and
        G. Vasilopoulos\inst{\ref{oas}} \and
        J. Wilms\inst{\ref{uerl}} 
       } 

\titlerunning{The Be/X-ray binary pulsar \src in the LMC}
\authorrunning{Maitra et al.}

\institute{
Max-Planck-Institut f{\"u}r extraterrestrische Physik, Gie{\ss}enbachstra{\ss}e 1, 85748 Garching, Germany\label{mpe}, \email{cmaitra@mpe.mpg.de}
\and
South African Astronomical Observatory, PO Box 9, Observatory Rd, Observatory 7935, South Africa\label{salt}
\and
Department of Astronomy, University of Cape Town, Private Bag X3, Rondebosch 7701, South Africa\label{uct}
\and
Astronomical Observatory, University of Warsaw, Al. Ujazdowskie 4, 00-478, 
Warszawa, Poland\label{uw}
\and
Institut f{\"u}r Astronomie und Astrophysik, Sand 1, 72076 T{\"u}bingen, Germany\label{iaat}
\and
Dr. Karl Remeis-Observatory and Erlangen Centre for Astroparticle Physics, Friedrich-Alexander-Universit\"t Erlangen-N\"urnberg, Sternwartstr. 7, 96049 Bamberg, Germany\label{uerl}
\and
Department of Physics and Astronomy, Howard University,
Washington, DC 20059, USA\label{howard}
\and
CRESST/Code 661 Astroparticle Physics Laboratory, NASA Goddard Space Flight Center, Greenbelt Rd., MD 20771, USA\label{gsfc}
\and
Universities Space Research Association,  Huntsville, US\label{usra}
\and
Universit\'e de Strasbourg, CNRS, Observatoire astronomique de Strasbourg, UMR 7550, F-67000 Strasbourg, France\label{oas}
}

\date{Received ... / Accepted ...}

\abstract
   {The Magellanic Clouds are our nearest star-forming galaxies. While the  population of high-mass X-ray binaries (HMXBs) in the Small Magellanic Cloud  (SMC) is relatively well studied, our knowledge about the Large Magellanic Cloud (LMC) is far from complete given its large angular extent and insufficient coverage with X-ray observations.}
   {We conducted a search for new HMXBs in the LMC using data from \ero, the soft X-ray instrument on board the Spektrum-Roentgen-Gamma (\srg) satellite.}
   {After confirming the nature of \src as a hard X-ray source positionally coincident with an early-type star, we followed it up with optical spectroscopic observations from South African Large Telescope (SALT) and a  dedicated \nus observation.}
   {We study the broadband timing and spectral behaviour of the newly discovered HMXB \src through \ero, \swift and \nus data in X-rays and  the Optical Gravitational Lensing Experiment (OGLE) and SALT RSS data at  optical wavelength. We report on detection of the spin period at 1412\,s and suggest an orbital period of the system of $\sim$151\,days, and thereby establish \src as an accreting pulsar. Further, through optical spectroscopic observations and the existence of \Halpha emission the source is identified as a Be X-ray binary pulsar in the LMC. We also investigate the variability of the source in the optical and X-ray regime over the last decades and provide estimates on the possible magnetic field strength of the neutron star.}
   {}

\keywords{galaxies: individual: LMC --
          X-rays: binaries --
          stars: emission-line, Be -- 
          stars: neutron
          pulsars: individual: \src
         }

\maketitle   


\section{Introduction}
\label{sec:intro}

The Magellanic Clouds are our nearest star-forming galaxies with current global star formation rates (SFR) of 0.021--0.05\,$M_{\odot}\,\mathrm{yr}^{-1}$ and 0.068--0.161\,$M_{\odot}\,\mathrm{yr}^{-1}$ for the Small Magellanic Cloud (SMC) and the Large Magellanic Cloud (LMC), respectively \citep{2018MNRAS.480.2743F}. The SMC especially is known to host a large population of Be X-ray binaries (BeXRBs), owing to a peak in the star formation activity between 25--40\,Myr ago \citep[][]{2016A&A...586A..81H,2016MNRAS.459..528A}. The high-mass X-ray binary (HMXB) population in the LMC is attributed to a star formation event at a slightly earlier epoch, and the formation efficiency of HMXBs in the LMC is also indicated to be lower in comparison to the SMC \citep[][]{2016MNRAS.459..528A}. The above information is, however, far from complete given the large angular extent and insufficient coverage of the LMC by X-ray observations, which implies that only part of the XRB population is known up to now.  

Our observational knowledge of the X-ray population of the Magellanic Clouds improved drastically with the launch of \ero, the soft X-ray instrument on board the Spektrum-Roentgen-Gamma (\srg) mission \citep{2021A&A...647A...1P} which surveyed the X-ray sky in an energy range of 0.2--8\,keV between December 2019 and February 2022. This has led to the discovery of a number of new HMXBs, as well as detection of pulsations from previous candidate HMXBs thus confirming their nature as a neutron star X-ray binary system. See for example \citet{2020ATel13609....1H,2020ATel13610....1M,2020ATel13650....1M,2021A&A...647A...8M,2022A&A...661A..25H,2022A&A...661A..20C}, and \citet{2021ATel15133....1H}.

\src was discovered during the course of the first \ero all-sky survey (eRASS1) as a new BeXRB in the LMC \citep{2020ATel13610....1M}. Follow-up  with the \nus observatory through an unanticipated target of opportunity (ToO) observation revealed coherent pulsations from the source at 1412\,s and confirmed its nature as a neutron star BeXRB \citep{2020ATel13650....1M}.
In this paper we report the X-ray broadband and multi-wavelength characteristics of \src. 

We describe the 
X-ray observations of \src using \ero all-sky survey, \swift, and \nus data, 
and optical observations from  the Optical Gravitational Lensing Experiment \citep[OGLE,][]{2008AcA....58...69U,2015AcA....65....1U} and  Southern African Large Telescope \citep[SALT;][]{Buckley2006} in Sect.~\ref{sec:data}.
In Sect.~\ref{sec:analysis} we present the identification of the optical counterpart, search for X-ray variability and pulsations and describe the broadband timing and spectral analysis with the \ero and \nus data. We report on the optical long-term monitoring recorded by the OGLE (Sect.~\ref{sec:ogle-analysis}), and present the optical spectrum obtained with the SALT (Sect.~\ref{sec:SALT-analysis}). We discuss the results in Sect.~\ref{sec:discussion} and provide the conclusions in Sect.~\ref{sec:conclusion}.

\section{Observations and data reduction}
\label{sec:data}

\subsection{\ero}
\label{sec:ero}

\src was discovered as a new bright and hard X-ray source after it was scanned for several weeks during the first all-sky survey (eRASS1) performed by the \ero instrument on board the Russian/German Spektrum-Roentgen-Gamma (SRG) mission. Until the end of 2021, \ero scanned the source during five epochs in the energy range of 0.2--8\,keV, as summarised in Table~\ref{tabobsero}.
The source was detected in all the epochs. 
To analyse the data, we used the \ero Standard Analysis Software System \citep[\eSASS version {\tt eSASSusers\_201009,}][]{2022A&A...661A...1B}. 

We extracted \ero light curves and spectra using the \eSASS task \texttt{srctool} following \cite{2021A&A...647A...8M} and \cite{2022A&A...661A..25H}. The source events were extracted using a circular region of radius 35\arcsec\ and background events were extracted using a circle of a larger radius from a nearby source free region. Events from all valid pixel patterns (PATTERN=15) were selected.
For spectral analysis we combined the data from the five telescope modules (TM) with on-chip filter cameras into a single spectrum (TM 1--4 and 6).  TM5 and TM7 were not used for spectroscopy because no reliable energy calibration is available so far due to an optical light leak which was discovered soon after the start of the CalPV phase \citep[][]{2021A&A...647A...1P}.
The spectra were binned to achieve a minimum of one count per spectral bin to allow the use of Cash statistic \citep{1979ApJ...228..939C}.  The light curves were created using all seven cameras and applying a cut in the fractional exposure of 0.15. Fractional exposure is a dimensionless quantity  which corresponds to the product of the fractional collecting area and the fractional temporal coverage overlapping with the time bin.

\subsection{\swift}
\label{sec:swift}

After its discovery with \ero, \src was followed up using the X-ray telescope \citep[XRT; ][]{2004ApJ...611.1005G, 2005SSRv..120..165B} on the Neil Gehrels Swift observatory on 2020-03-30 (MJD 58938.57--58938.98, OBSID 00013298003) with a 2.3\,ks exposure. 
The \swift/XRT data were analysed following the \swift data analysis guide\footnote{\url{http://www.swift.ac.uk/analysis/xrt/}} \citep{2007A&A...469..379E}. Source detection was performed using a sliding-cell detection algorithm implemented by \texttt{XIMAGE} and \texttt{sosta}\footnote{\url{https://heasarc.gsfc.nasa.gov/xanadu/ximage/ximage.html}} using \texttt{HEASOFT} 6.28 and CALDB version 20200724. 
The source was detected with an XRT count rate of $(2.42\pm0.35) \times10^{-2}$\,\cts\ in the 0.3--10.0\,keV band. 
The \swift/XRT spectrum could be described by an absorbed power-law with photon index $\Gamma=0.4^{+1.4}_{-0.4}$ and $\nh=4^{+1.9}_{-0.4}$\,\ohcm{21}. 
The observed 0.3--10\,keV flux of 2.5\ergcm{-12} corresponds to an unabsorbed luminosity of  7\ergs{35}, for a source distance of 50\,kpc which is assumed throughout the remainder of this paper \citep[][]{2013Natur.495...76P,2019Natur.567..200P,2019ApJ...876...85R}.

\subsection{\nus}

Following its discovery, \src was also followed up by an unanticipated ToO observation by \nus  \citep[][]{2013ApJ...770..103H} on 2020-04-08 (MJD 58947, Obsid 90601312002, net exposure 63\,ks). 
\nus consists of two independent focal plane modules FPMA and FPMB operating in the energy range of 3--78\,keV. The data were processed from both modules using the standard \texttt{NuSTARDAS} 
software (version 1.8.0 of \texttt{HEASOFT} 6.28 and CALDB version 20200425) to create barycenter-corrected light curves, spectra, response matrices,
and effective area files. The source events were extracted using a circular region of radius 20\arcsec\ and background events were extracted using an annulus region with an inner and outer radii of 43\arcsec\ and 65\arcsec\ respectively.
The observation was free from stray light contamination.
We extracted a background-subtracted light curve in different energy bands with a time resolution of 100\,ms, combining the events from both \nus modules.

\begin{table*} 
\centering
\caption{\ero Observations of \src} 
\label{tabobsero} 
\begin{tabular}{ccccccc} 
\hline\hline\noalign{\smallskip}
Epoch & \ero survey\tablefootmark{a} & Obs. time & Net exposure\tablefootmark{b} & count rate\tablefootmark{c} & Var\tablefootmark{d} & S\tablefootmark{d}\\
      &            &   T$_{start}$ -- T$_{stop}$ (UTC) & ks & \cts & & \\
\noalign{\smallskip}\hline\noalign{\smallskip}
 1 & eRASS0-eRASS1 & 2019-12-08 16:01:46 -- 2019-12-20 23:32:02 & 0.9 &  0.28$\pm0.02$ & 15.0 & 3.2 \\ 
 2 & eRASS1-eRASS2 & 2020-05-30 00:57:55 -- 2020-06-14 20:57:56 & 1.3 &  0.18$\pm0.01$ & 6.3 & 2.6 \\
 3 & eRASS2-eRASS3 & 2020-12-05 22:31:57 -- 2020-12-23 01:32:04 & 1.2 &  0.19$\pm0.01$ & 5.1 & 2.4 \\
 4 & eRASS3-eRASS4 & 2021-06-02 09:58:00 -- 2021-06-20 09:58:09 & 0.8 &  0.10$\pm0.01$ & 4.9 & 2.3 \\
 5 & eRASS4-eRASS5 & 2021-12-09 11:32:03 -- 2021-12-27 03:32:06 & 1.3 &  0.16$\pm0.01$ & 2.9 & 1.9 \\
\noalign{\smallskip}\hline
\end{tabular} 
\tablefoot{
\tablefoottext{a}{Test-scans which were performed before the official start of the survey between 2019-12-08 and  2019-12-11 are designated as eRASS0.}
\tablefoottext{b}{Net exposure after correcting for vignetting and normalised to seven telescope modules. 
The lower exposure in epoch 4 is caused by only 4 active cameras (TM1, 2, 3, and 6) for most of the epoch.}
\tablefoottext{c}{Count rate in the 0.2--8\,keV energy band.}
\tablefoottext{d}{Variability Var of the light curve within the given epoch and significance S of the stated variability. 
For parameter definitions see Sect.\,\ref{sec:time-analysis}.}
}
\end{table*}

\subsection{OGLE}
\label{sec:ogle}

The optical counterpart of \src was observed by OGLE, which started observations in 1992 \citep{1992AcA....42..253U} but interrupted since March 2020. 
\src was monitored regularly in the I (OGLE-ID LMC519.13.19023) and V (LMC519.13.v.33) filter bands during phase IV of the OGLE project \citep[][]{2015AcA....65....1U}. 
Images were taken in I (54\% of the measurements are separated by 1--2 days) and less frequently in V (54\% of the measurements every 1--4 days). The photometric magnitudes were calibrated to the standard VI system.

\subsection{SALT}
\label{sec:salt}
Optical spectroscopy of \src was undertaken on 2020-03-20 using the Robert Stobie Spectrograph (RSS) on the SALT under the SALT transient follow-up program. The PG0900 VPH grating was used, which covered the spectral region 3920--7000\,\AA\ at a resolution of 6.2\,\AA. A single 1200\,s exposure was obtained, starting at 18:33:47 UTC. The SALT pipeline was utilised to perform primary reductions comprising overscan corrections, bias subtraction, gain correction and amplifier cross-talk corrections \citep[][]{2010SPIE.7737E..25C}. The remaining steps, which include wavelength calibration, background subtraction and extracting the 1D spectrum were executed using IRAF\footnote{https://iraf-community.github.io/}.

\section{Analysis}
\label{sec:analysis}
\subsection{X-ray position  and identification of optical counterpart}
\src was discovered during eRASS1 at a best-determined X-ray position (after astrometric correction) of $\alpha_\mathrm{J2000.0} = 05^\mathrm{h}29^\mathrm{m}14\fs68$ and $\delta_\mathrm{J2000.0} = -66\degr24\arcmin45\farcs2$ with a $1\sigma$ statistical uncertainty of 0\farcs6. The positional uncertainty is dominated by systematic astrometric uncertainties, and is typically of the order of 1\arcsec\ in the \ero\ survey data.
The count rate of the source as detected in each of the epochs in the energy range of 0.2--8\,keV (after correcting for vignetting and a finite PSF aperture) is given in Table~\ref{tabobsero}. The hard X-ray spectrum for a point-like source makes it an ideal candidate for a HMXB. Additionally, there were no AGN or background galaxies from known catalogs coincident with the X-ray source position. Therefore, we searched for the most-likely optical counterpart in the
Magellanic Clouds Photometric Survey \citep{2004AJ....128.1606Z}
and the Two Micron All Sky Survey \citep[2MASS,][]{2003yCat.2246....0C} and identified a unique counterpart within the X-ray error circle (2MASS\,05291423-6624440).
The magnitudes and colours are consistent with an early type star, suggesting \src is a BeXRB in the LMC. The \gaia counterpart (Gaia\,DR3\,4660325634049868800) also supports the scenario indicating an effective temperature of 24233-24571\,K (1$\sigma$ range).
Table\,\ref{tab:optical} details some properties of the optical counterparts.
\begin{table*}
\centering
\caption{Optical counterpart of \src}
\label{tab:optical}
\begingroup
\begin{tabular}{ccccccccc}
\hline\hline\noalign{\smallskip}
\multicolumn{1}{c}{V\tablefootmark{a}} &
\multicolumn{1}{c}{Q\tablefootmark{a,b}} &
\multicolumn{1}{c}{2MASS} &
\multicolumn{1}{c}{J\tablefootmark{c}} &
\multicolumn{1}{c}{H\tablefootmark{c}} &
\multicolumn{1}{c}{K$_{\rm s}$\tablefootmark{c}} &
\multicolumn{1}{c}{R.A.} &
\multicolumn{1}{c}{Dec.} &
\multicolumn{1}{c}{D\tablefootmark{e}} \\
\multicolumn{6}{c}{} &
\multicolumn{2}{c}{(J2000)\tablefootmark{d}} &
\multicolumn{1}{c}{} \\
\multicolumn{1}{c}{(mag)} &
\multicolumn{1}{c}{(mag)} &
\multicolumn{1}{c}{} &
\multicolumn{1}{c}{(mag)} &
\multicolumn{1}{c}{(mag)} &
\multicolumn{1}{c}{(mag)} &
\multicolumn{1}{c}{(h m s)} &
\multicolumn{1}{c}{(\degr\ \arcmin\ \arcsec)} &
\multicolumn{1}{c}{(\arcsec)} \\
\noalign{\smallskip}\hline\noalign{\smallskip}
14.42 & -0.55 & 05291423$-$6624440 & 14.66 & 14.85 & 14.59 & 05 29 14.25 & -66 24 44.02 & 2.8 \\
\noalign{\smallskip}\hline
\end{tabular}
\endgroup
\tablefoot{
\tablefoottext{a}{ V magnitude and reddening-free parameter, defined as $Q = \mathrm{U}-\mathrm{B} - 0.72(\mathrm{B}-\mathrm{V})$ are taken from \citet{2004AJ....128.1606Z}}
\tablefoottext{b}{For the distribution of the Q parameter of BeXRBs in the SMC see \citet{2016A&A...586A..81H}.}
\tablefoottext{c}{J, H, K$_{\rm s}$ magnitudes are taken from the 2MASS counterpart.}
\tablefoottext{d}{Position of the optical counterpart from Gaia DR3 \citep[see][]{2022arXiv220605989B}.}
\tablefoottext{e}{Angular distance between \ero and \gaia position.}
}
\end{table*}


\subsection{Timing analysis: Searching for variability and pulsations}
\label{sec:time-analysis}
\begin{figure}
\centering
\resizebox{\hsize}{!}{\includegraphics{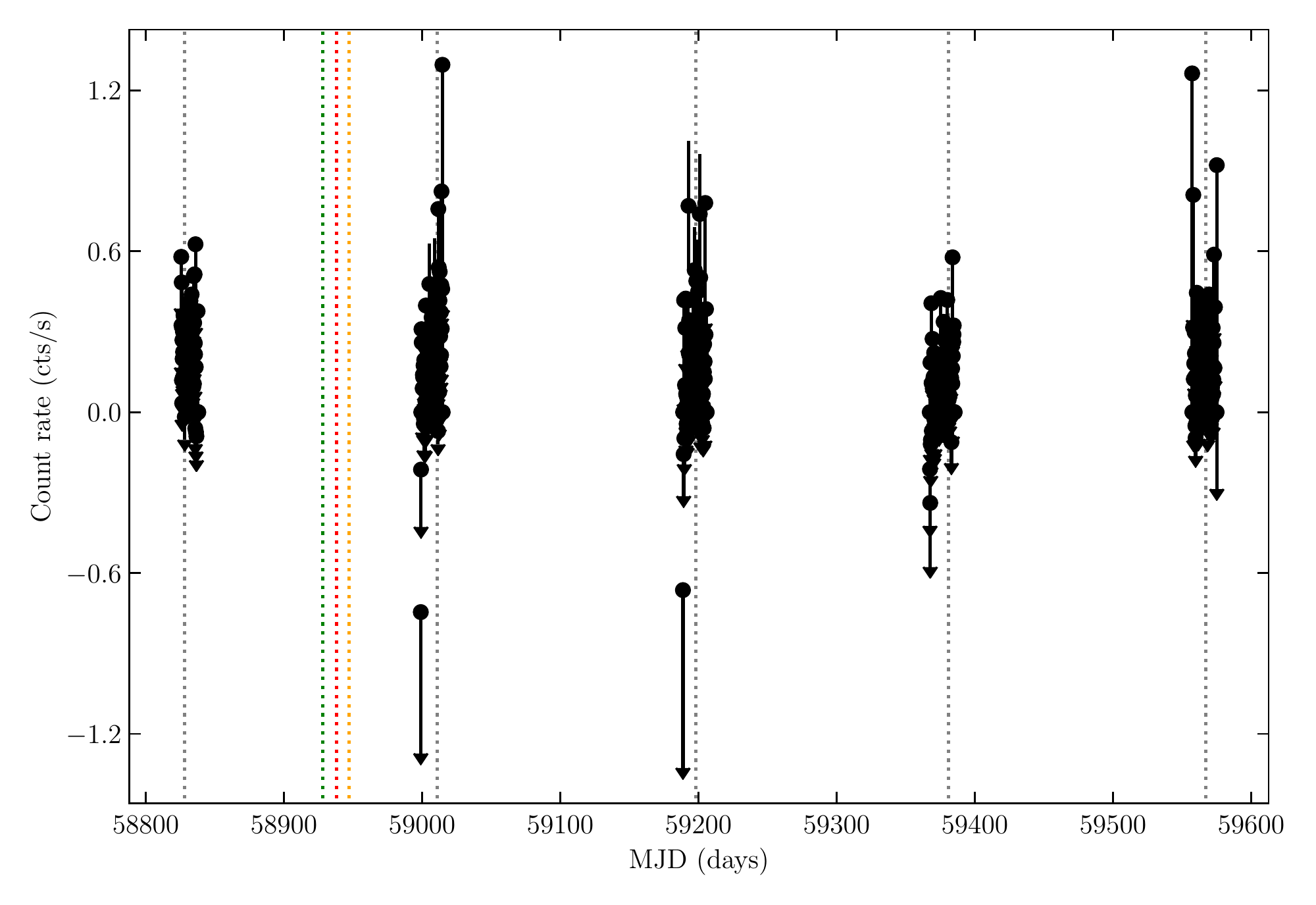}}

  \caption{\ero light curve of \src in the five epochs from eRASS0 to beginning of eRASS5. Each point represents a single scan over the source.
  The green, red and orange dotted lines indicate the date of the SALT, \swift and \nus follow-up observations respectively. The black dotted lines indicate the formal transitions between different eRASSs.  Error bars in $y$-direction show $1\sigma$ uncertainties assuming Poisson statistic. Count rates of bins with less than 10 counts are plotted as 1$\sigma$ upper limits.
  }
  \label{fig:eROLC_full}
\end{figure}

The \ero surveys scanned the whole sky in great circles which intercepted at the ecliptic poles. The scanning period is 4 hours and given the $1^\circ$ field of view, a typical scan lasted for up to 40\,s, with scan separations of 4\,hours. In the case of a source located in the direction of the LMC, which is situated close to the South Ecliptic Pole, it is scanned more often than at the ecliptic plane, i.e., for up to several weeks per eRASS. This provides a unique opportunity to probe the variability of HMXBs on timescales of few tens of seconds to several weeks. Each \ero survey (eRASSn) is repeated after six months which then allows a source to be monitored also on time scales of years. We present here the  \ero data of \src from the first four complete all-sky surveys (eRASS1, 2, 3, and 4) and the beginning of eRASS5.
Figure~\ref{fig:eROLC_full} shows the 0.2--8\,keV \ero light curve of \src as it was scanned during each eRASS at various epochs described in Sect.~\ref{sec:ero}. The mean count rate of \src during each of these epochs are given in Table~\ref{tabobsero}. As is evident, the source was brightest during epoch 1 and exhibited comparable count rates during the subsequent epochs.

\begin{figure*}[ht]
\centering
\begin{tabular}{@{}c@{}c@{}}
    \includegraphics*[width=\columnwidth]{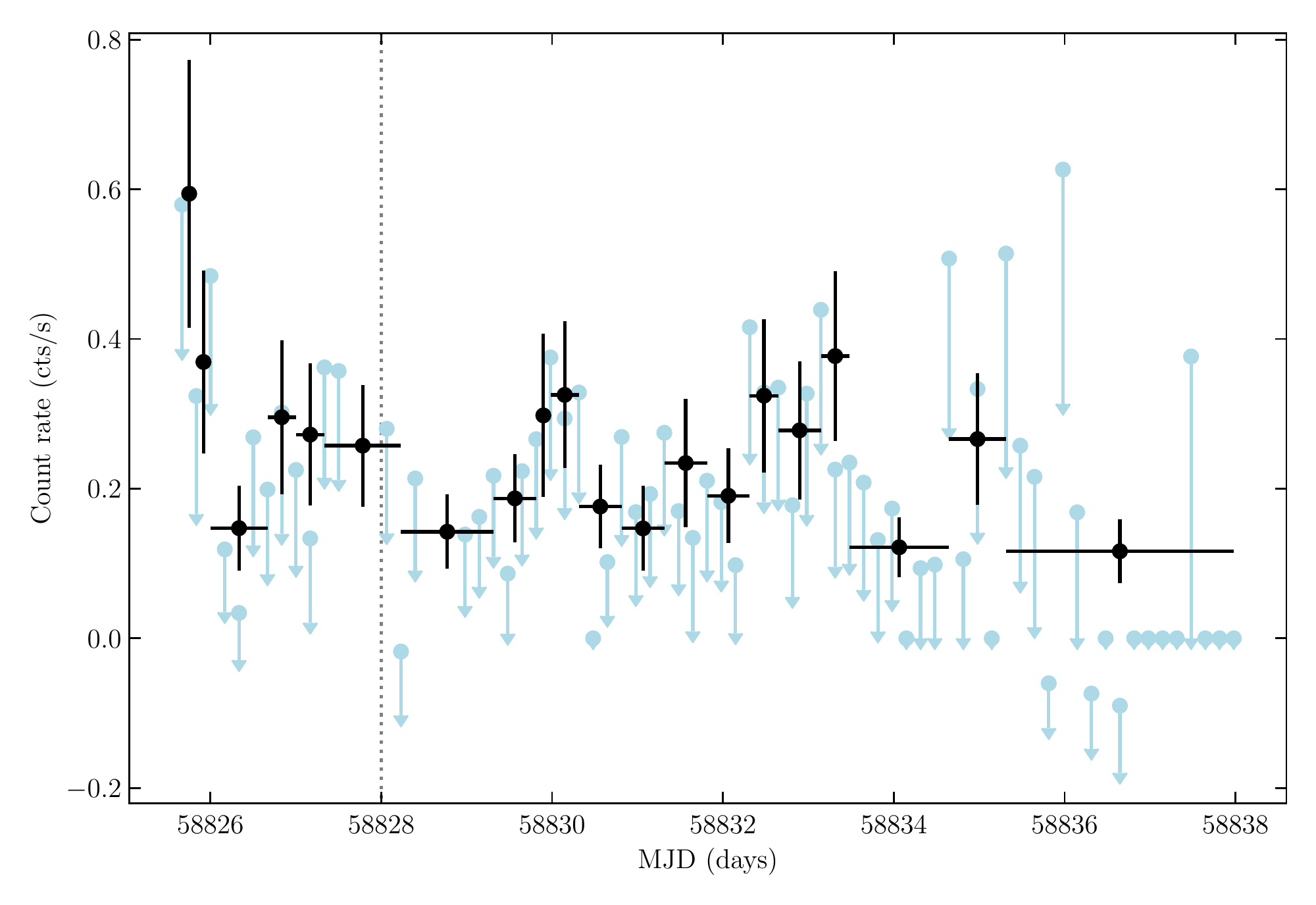} & 
    \includegraphics*[width=\columnwidth]{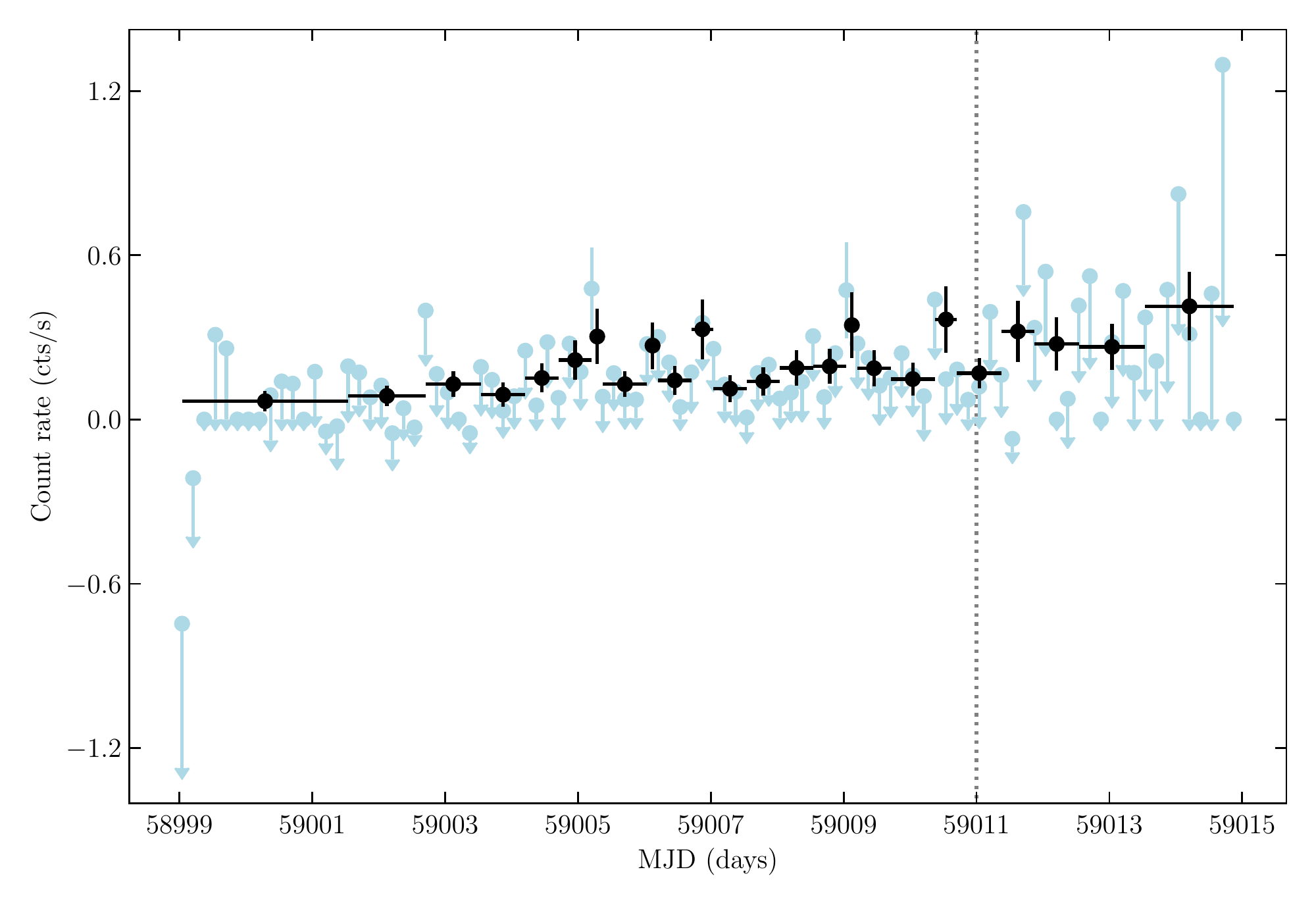} \\
    \includegraphics*[width=\columnwidth]{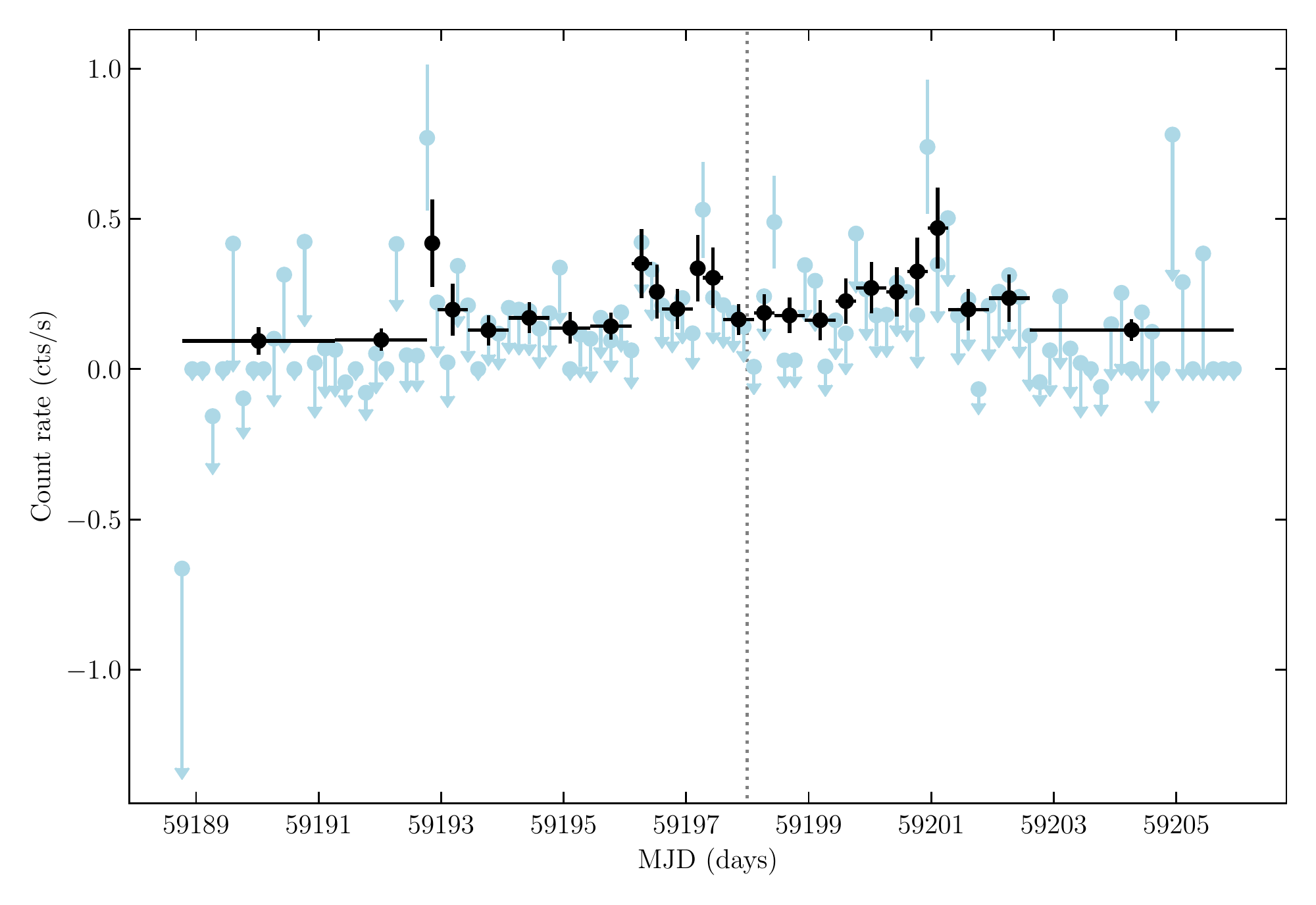} & 
    \includegraphics*[width=\columnwidth]{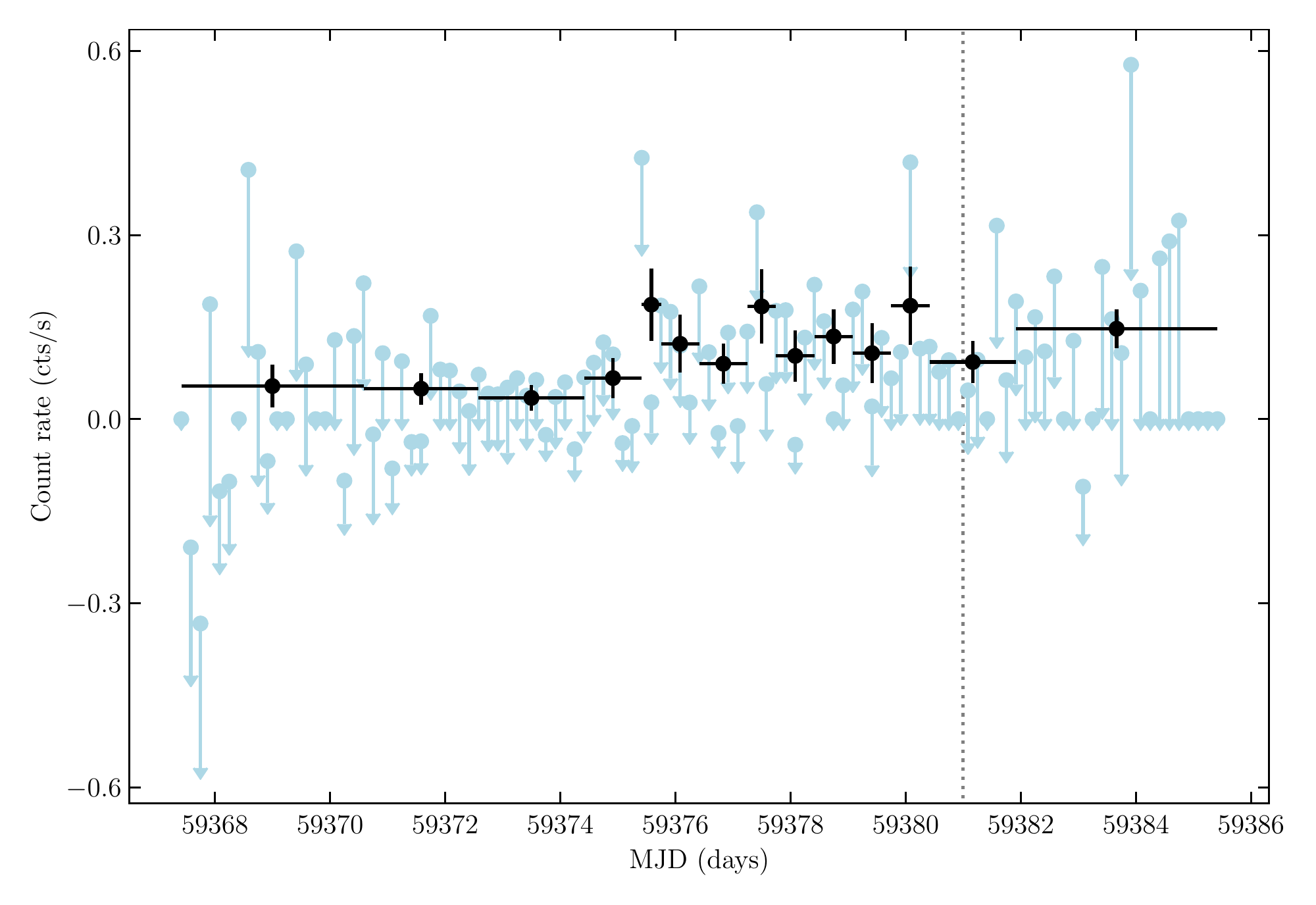} \\
    \multicolumn{2}{c}{\includegraphics*[width=\columnwidth]{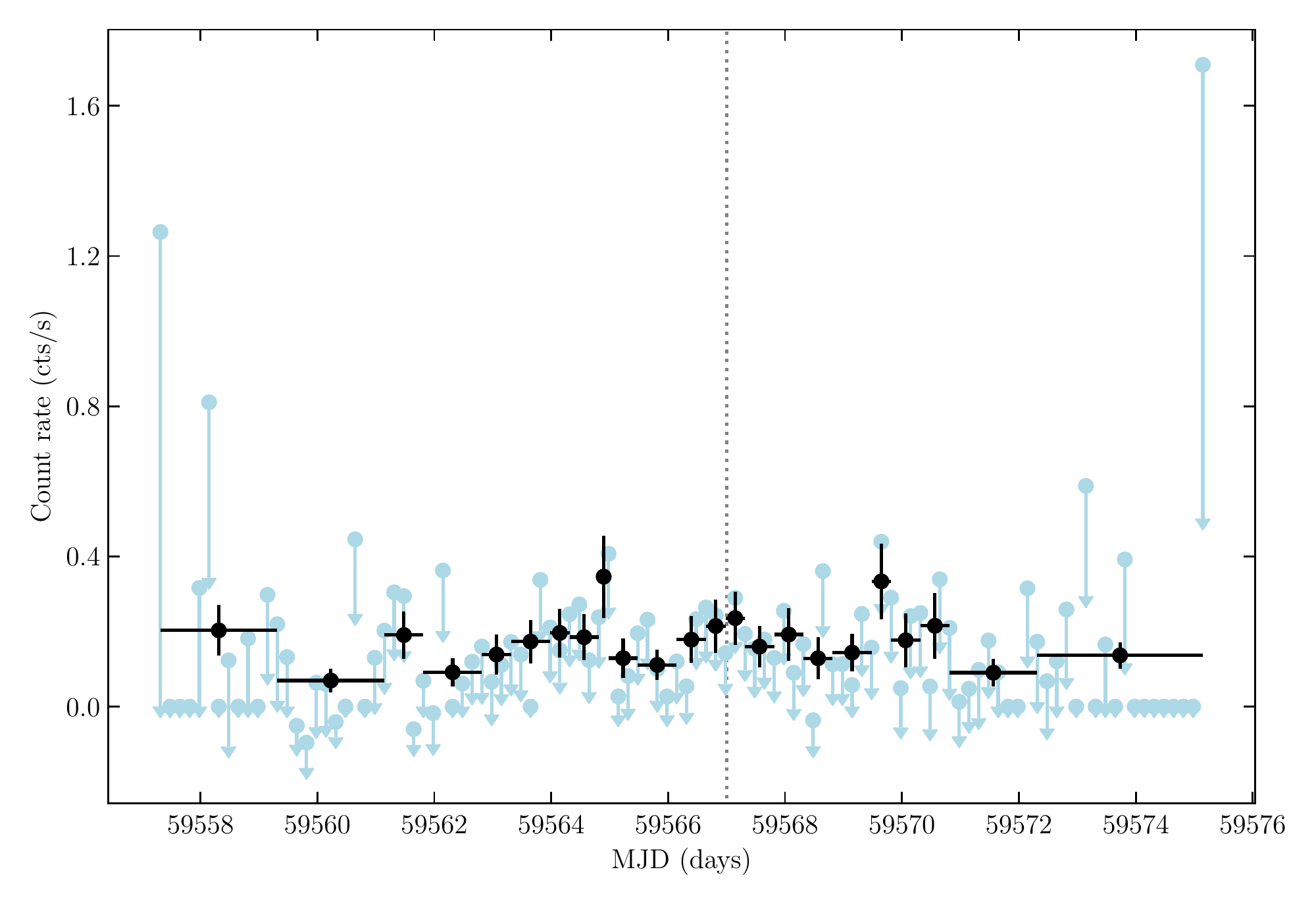}}
\end{tabular}
\caption{\ero light curves of \src, where the blue points represent a single scan and the black points are rebinned to a minimum number of 10\,counts per bin for better statistical treatment. Error bars in $x$-direction indicate time bins, error bars in $y$-direction show $1\sigma$ uncertainties assuming Poisson statistic. Count rates of bins with less than 10 counts are plotted as 1$\sigma$ upper limits. The black dotted lines indicate the formal transitions between different eRASSs. In light blue the original observation pattern obtained by \ero can be seen.  Shown are light curves for epochs~1 (top left), 2 (top right), 3 (centre left), 4 (centre right), and~5 (bottom) as defined in Table\,\ref{tabobsero}.}
\label{fig:eRO_LC}
\end{figure*}

We investigated variability within single survey epochs similar to \cite{2008A&A...480..599S} and \cite{2013A&A...558A...3S} defining the variability $Var$ and significance of the difference $S$ as
\begin{equation*}
    Var = \frac{F_{max}}{F_{min}} \\
    S = \frac{F_{max} - F_{min}}{\sqrt{\sigma_{max}^2 + \sigma_{min}^2}},
\end{equation*}
where $F_{max}$ and $\sigma_{max}$ define the source flux (background subtracted) and 1$\sigma$ confidence interval of the time bin for which $F - \sigma$ is maximal and analogously $F_{min}$ and $\sigma_{min}$ of the time bin for which $F + \sigma$ is minimal. In order to get a better statistical treatment we rebinned the \ero light curves of \src by summing up original bins until there are at least 10 photon counts per time bin. The resulting light curves are displayed as black points in Fig.~\ref{fig:eRO_LC}. Using these newly defined bins we find the largest variability of $Var=15.0$ in epoch 1 with a significance of $S=3.2$. Values for $Var$ and $S$ for all epochs can be found in Table~\ref{tabobsero}. 

\begin{figure}
\begin{center}
 \resizebox{\hsize}{!}{\includegraphics{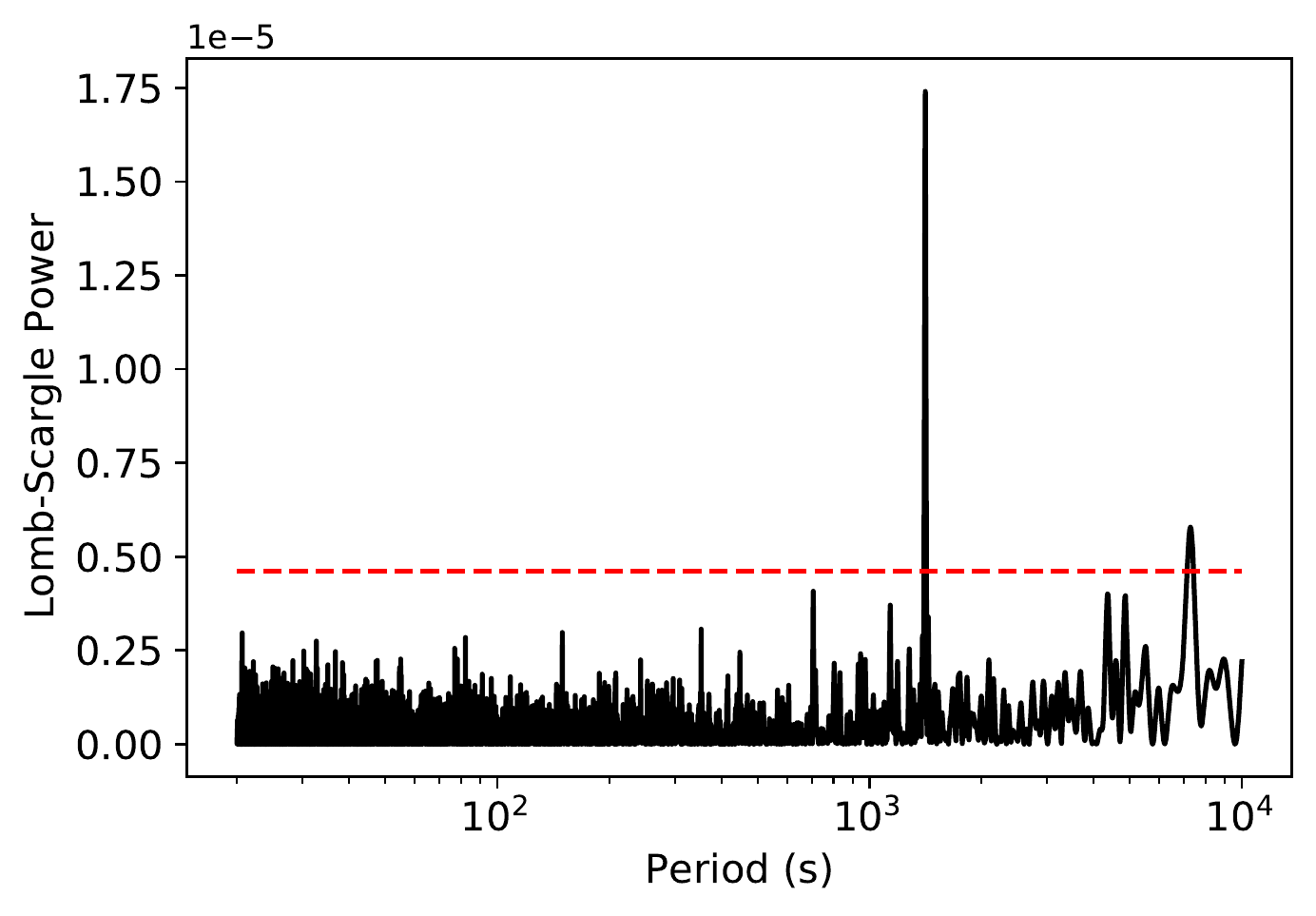}}
\end{center}
  \caption{
    Lomb-Scargle periodogram of \src obtained from the combined \nus data (3.0--30.0 keV). 
    Pulsations are clearly detected with of period of 1412\,s.
    The red dashed line marks the 99.73\% confidence level obtained by block bootstrapping method.}
  \label{fig:PNpower}
\end{figure}

\begin{figure}
  \centering
  \resizebox{0.9\hsize}{!}{\includegraphics[]{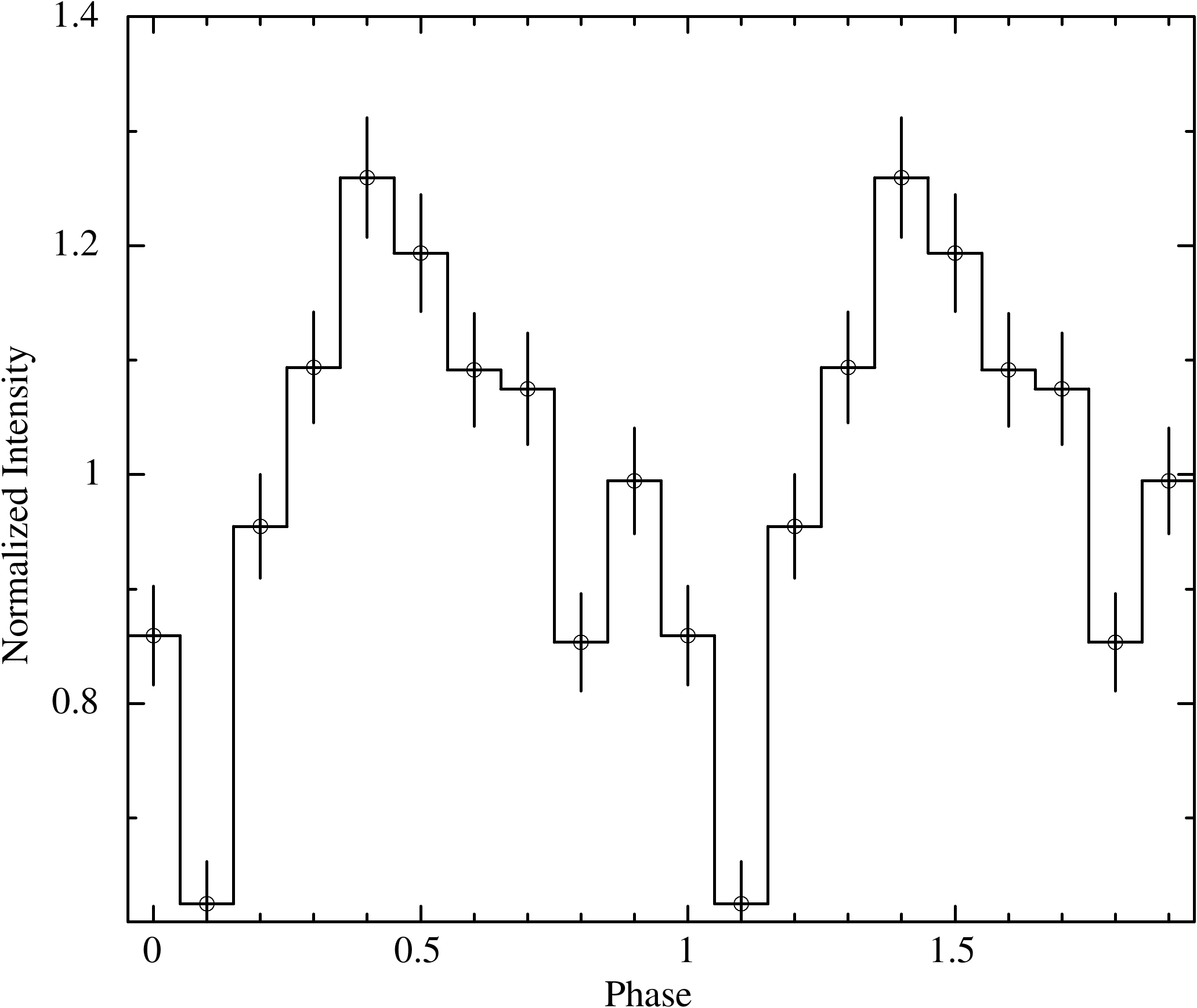}}
  \caption{
    Background subtracted \nus light curve folded with 1411.8\,s showing the pulse profile of \src in the energy band of 3-78\,keV. The pulse profile is normalised to the mean count rate of  0.14\,\cts.
  }
  \label{fig:pp}
\end{figure}

\begin{figure}
  \centering
  \resizebox{0.7\hsize}{!}{\includegraphics[]{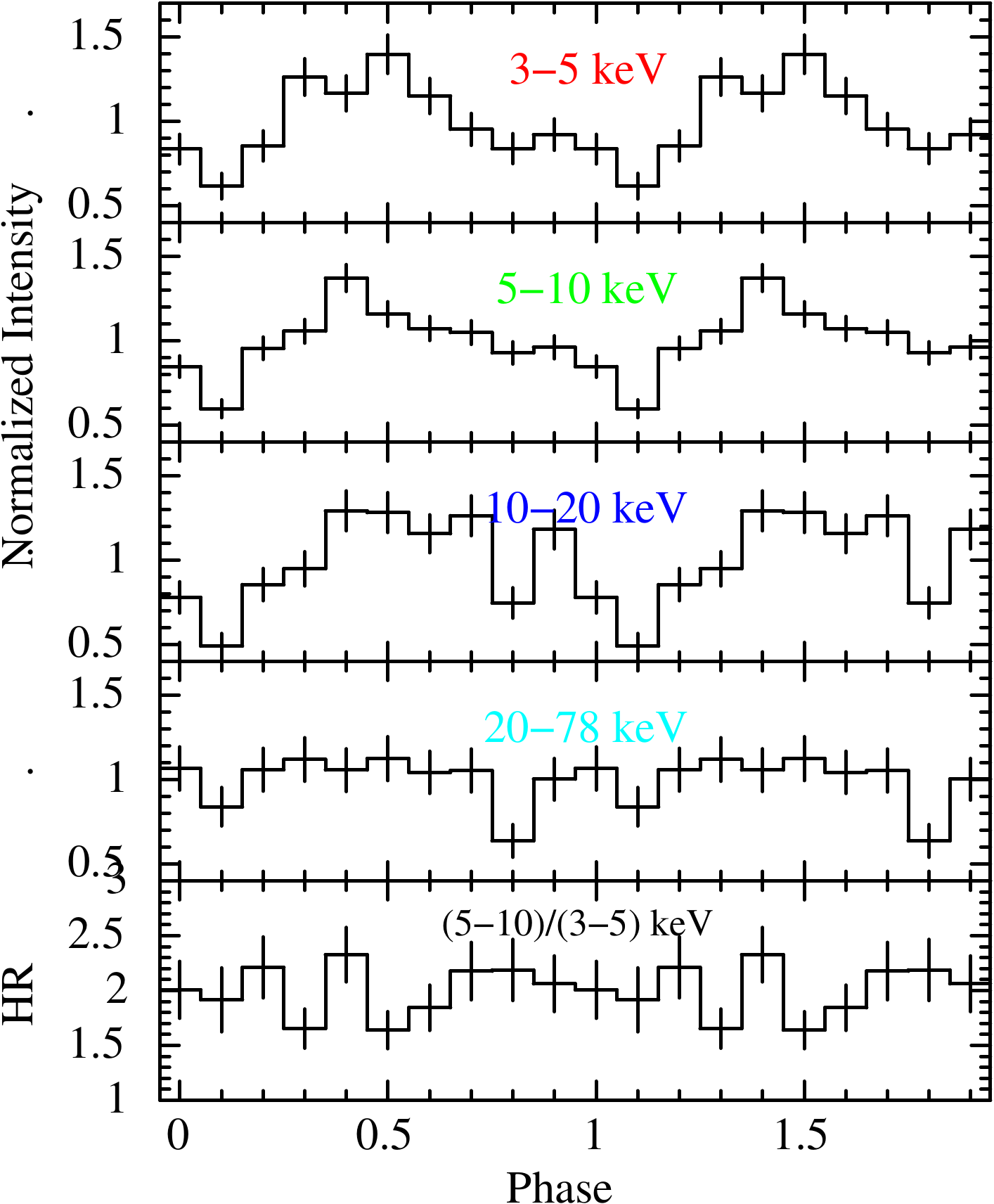}}
  \caption{
    Background subtracted \nus light curve folded with 1411.8\,s showing the pulse profile of \src in the energy ranges and the Hardness ratio (HR) defined in the figure. The pulse profiles are normalised to the mean count rate of  0.032\,\cts,  0.064\,\cts, 0.028\,\cts, and 0.020\,\cts\, respectively.
  }
  \label{fig:pp_e}
\end{figure}

Using the data from the dedicated \nus observation, we searched for a periodic signal using a Lomb-Scargle (LS) periodogram analysis \citep{1976Ap&SS..39..447L,1982ApJ...263..835S} in the range of 1--5000\,s, which covers typical periods seen from HMXB pulsars \citep[][]{2012MmSAI..83..230C}. A very strong signal is detected at $\sim$1412\,s (see Fig.~\ref{fig:PNpower}), indicating the spin period of the neutron star. To determine the spin periods more precisely, an epoch-folding technique was used \citep{1987A&A...180..275L}. The best-determined spin period and the uncertainty are given by 1411.8$\pm$1.7\,s. An additional peak is visible at $\sim$7000\,s corresponding to approximately five times the fundamental frequency. However it is to be noted that red-noise is commonly observed in accreting sources and can often appear periodic \citep[e.g.][]{1981ApJS...45....1S}. Therefore the false-alarm probability of the peaks in the periodogram at lower frequencies can be underestimated \citep[][]{2008ApJS..177..181L}. 
To estimate the uncertainty on the spin period, we applied a block-bootstrapping method, similarly to \cite{Gotthelf1999}, and \citet{2022A&A...661A..20C} by generating a set of 10\,000 light curves. The background subtracted \nus light curve obtained by combining both modules and folded with the best-obtained period is shown in Fig.~\ref{fig:pp}. The pulsed fraction, computed by integrating over the 3-78\,keV energy band and defined as the ratio of $(I_{max}-I_{min})/(I_{max}+I_{min})$, is $31\pm$3\%.
We further investigated a possible dependence of the pulse profile with energy by extracting light curves in energy bands with comparable statistical quality. Figure~\ref{fig:pp_e} shows that the pulse shape morphs from a dip like feature around phase bin $\sim$0.7 to a broad shoulder-like structure between the energy range of 3--5\,keV to 5--10\,keV. The pulsed fraction in the energy range of 3--5\,keV, 5--10\,keV,10--20\,keV and 20--78\,keV is  $44\pm$8\%, $34\pm$6\% $47\pm$8\% and $28\pm$14\% respectively. Pulsations are not significantly detected above $>20$\,keV, which is reflected in the low pulsed fraction of the source and large fractional errors and a lower signal to noise ratio in the highest energy band.

\subsection{Long-term X-ray variability}
\label{sec:long_LC}

\begin{figure}
   \centering
   \resizebox{1.1\hsize}{!}{\includegraphics{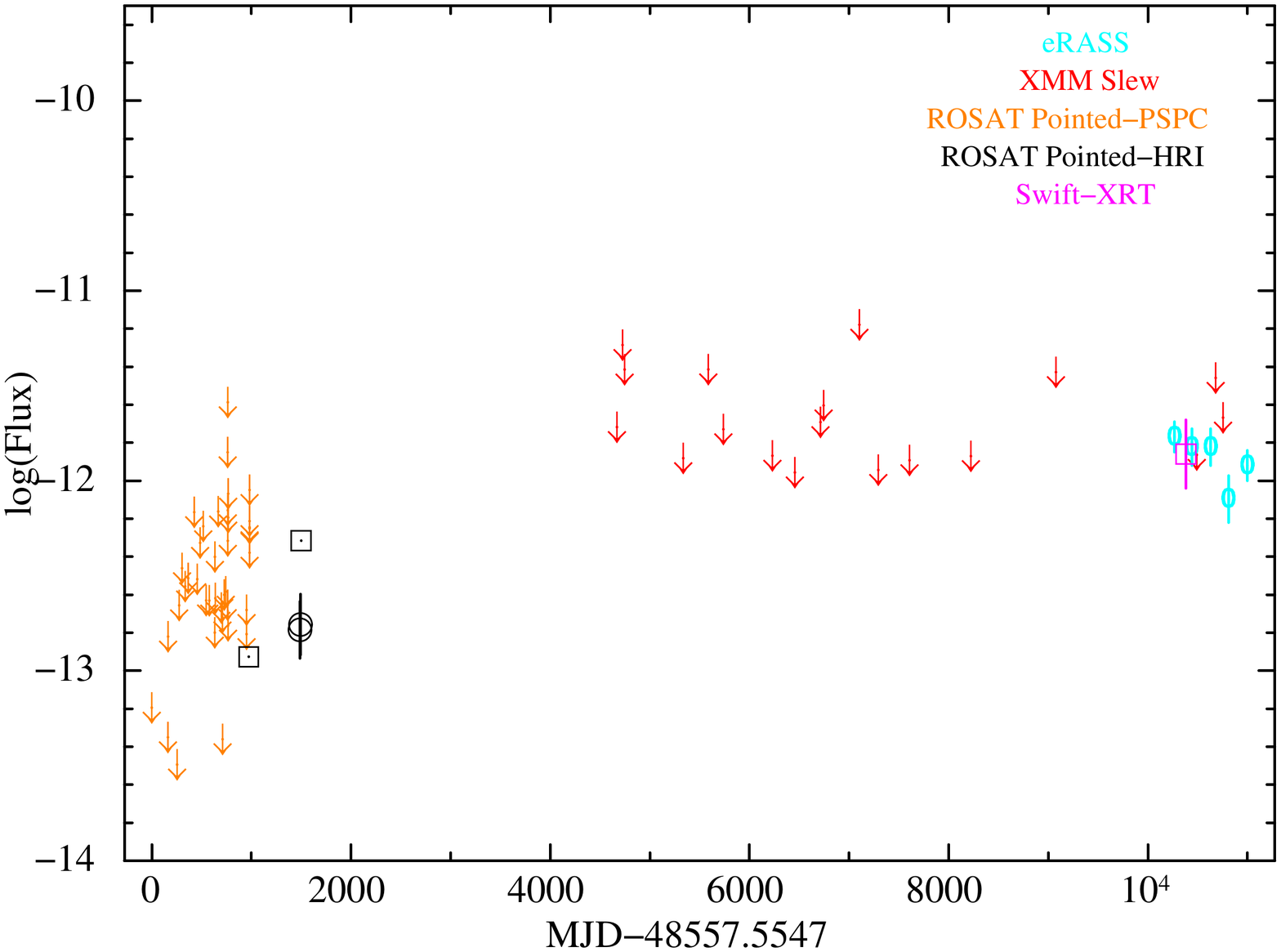}}
   \caption{Long-term X-ray light curve of \src in the energy range of 0.2--2\,keV. Circles and ellipses represent measured flux in units of erg cm$^{-2}$ s$^{-1}$, while upper limits are shown as arrows. 
   \ROSAT PSPC, \ROSAT HRI, \xmm (slew), \ero and \swift fluxes are shown in orange, black, blue and cyan and magenta respectively.
   Flux values except \ero measurements are derived using the High-Energy Lightcurve Generator.}
    \label{fig:long_LC}             
\end{figure}

Figure~\ref{fig:long_LC} shows the long-term X-ray light curve of \src in the energy range of 0.2--2\,keV. Circles and ellipses represent measured flux in units of \uergcm, while upper limits are shown as arrows. The fluxes are derived from  the \xmm slew, \ROSAT PSPC and HRI observations using the ``High-Energy Lightcurve Generator'' \citep[ULS\footnote{\url{http://xmmuls.esac.esa.int/upperlimitserver}};][]{2022A&C....3800531S,2022A&C....3800529K} 
and are derived assuming the spectral parameters from the simultaneous fitting of the \ero spectra as given in Table~\ref{tabobsero}. The source was detected twice around MJD~50047 in two \ROSAT HRI observations with a factor of $<10$ lower flux than during the eRASS scans. The limit is, however, quite uncertain due to the complex nature of the spectrum at low energies (see Sect.~\ref{sec:spec-analysis}) and foreground absorption.

\subsection{Simultaneous spectral fitting of the \ero spectra}
\label{sec:spec-analysis}

\begin{figure}
\centering

\resizebox{\hsize}{!}{\includegraphics{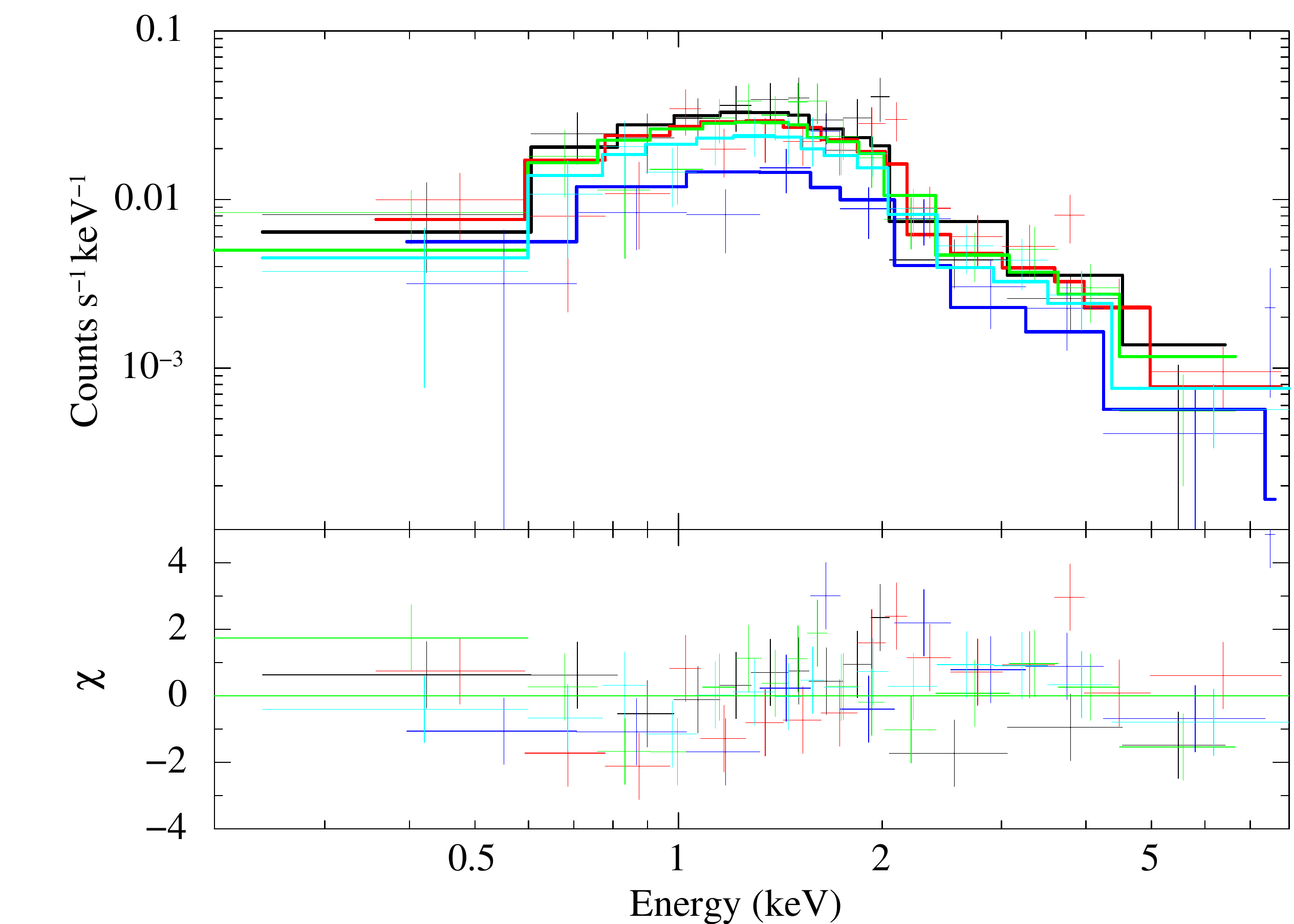}}
  \caption{Simultaneous spectral fit of \src using \ero spectra of the different epochs defined in Table \ref{tabobsero}, namely epoch 1 (black), epoch 2 (red), epoch 3 (green), epoch 4 (blue) and epoch 5 (turquoise) together with the best-fit model as histograms. The spectrum for each eRASS is extracted by combining data from the TMs with on-chip filter.
The residuals are plotted in the lower panels.}

  \label{fig:eRO_spec}
\end{figure}

The spectra of \src extracted from each of the epochs were simultaneously fit to constrain the average spectral model as well to investigate possible variability between each eRASSn. 
All spectra were binned to achieve a minimum of one count per spectral bin and C-statistic was used because of the low statistics. Spectral analysis was performed using \xspec v12.11.1k \citep{1996ASPC..101...17A}. 
To account for the photo-electric absorption by the Galactic interstellar gas we used \texttt{tbabs} in \xspec with ISM
abundances following \citet{2000ApJ...542..914W} with atomic cross sections from \citet{1996ApJ...465..487V}. 
The Galactic column density of $6\times10^{20}$ cm$^{-2}$ was taken from  \citet{1990ARA&A..28..215D} and fixed in the fits.
For the absorption along the line of sight (which includes the component through the LMC as well as local to the source), when required we included a second absorption component with elemental abundances fixed at 0.49 solar \citep{2002A&A...396...53R,1998AJ....115..605L}. We left the column density for this component free in the fit. Errors were estimated at 90\% confidence intervals. 
A simple absorbed power-law model provided an adequate description of the spectrum.
The X-ray spectrum of \src was modeled using a simultaneous fit of the data from epochs 1, 2, 3, 4 and 5 with only the power-law normalisations left free (it was not possible to investigate possible changes in the power-law index in the current work).  The potential additional absorption component from within the LMC could not be constrained and did not improve the fit statistic which is the reason we only included the Galactic absorption in our final fit. 
The parameters of the best-fit model are listed in Table~\ref{tcb:eRO_spec} and the spectra and best-fit model are plotted in Fig.~\ref{fig:eRO_spec}.

\begin{table*}
    \caption{Best fit parameters of \src obtained from the simultaneous fitting of epochs 1-5 spectra (top) and epoch 1 and \nus spectrum (bottom).}
    \centering
    \begin{tabular}{ccccc}
    \hline\hline\noalign{\smallskip}
        Epoch & \ero survey & Power law & \fx$^{a}$ & \lx$^{b}$ \\
        -- & -- & index & \oergcm{-12} & \oergs{35} \\
    \noalign{\smallskip}\hline\noalign{\smallskip}
        1 & eRASS0-eRASS1 &  & 1.7$^{+0.4}_{-0.3}$ & 5.1$^{+1.1}_{-1.0}$\\ \noalign{\smallskip}
        2 & eRASS1-eRASS2 &  & 1.5$\pm0.3$ & 4.5$^{+1.0}_{-0.8}$\\ \noalign{\smallskip}
        3 & eRASS2-eRASS3 & $0.28\pm0.13$ & 1.5$\pm0.3$ & 4.6$^{+1.0}_{-0.9}$\\ \noalign{\smallskip}
        4 & eRASS3-eRASS4 &  & 0.8$\pm0.2$ & 2.5$^{+0.7}_{-0.6}$\\ \noalign{\smallskip}
        5 & eRASS4-eRASS5 &  & 1.2$^{+0.3}_{-0.2}$ & 3.7$^{+0.9}_{-0.7}$\\ \noalign{\smallskip}
        \end{tabular}
        
       \begin{tabular}{ccccccc}
    \hline\noalign{\smallskip}
     eRASS + \nus   & Power law & cutoff energy   & \nh$_{\rm tbpcf}$ & CF & \fx$^{c}$    & \lx$^{b}$ \\
       --           & index     & keV             & \ohcm{21}         & -- & \oergcm{-12} &  \oergs{35} \\
    \noalign{\smallskip}\hline\noalign{\smallskip}
       --           & $0.97^{+0.29}_{-0.30}$ & 9.4$^{+3.7}_{-2.2}$ & 3.2$^{+2.3}_{-1.3}$ & 0.84$^{+0.09}_{-0.06}$ & 2.5$\pm0.2$ & 8.0$\pm0.7$ \\
    \noalign{\smallskip}\hline
    \end{tabular}
    \tablefoot{\tablefoottext{a}{Observed flux in the energy range of 0.2--8\,keV.}
    \tablefoottext{b}{Corresponding unabsorbed luminosity assuming a distance of 50\,kpc to \src.}
   \tablefoottext{c}{Observed flux in the energy range of 0.2--78\,keV.}
   The \nh\ is fixed to  the Galactic foreground value with solar abundances for the simultaneous eRASS spectral fit.}
    \label{tcb:eRO_spec}
\end{table*}


\subsection{Broadband spectrum of \src}

\begin{figure}
  \centering
  \resizebox{\hsize}{!}{\includegraphics[]{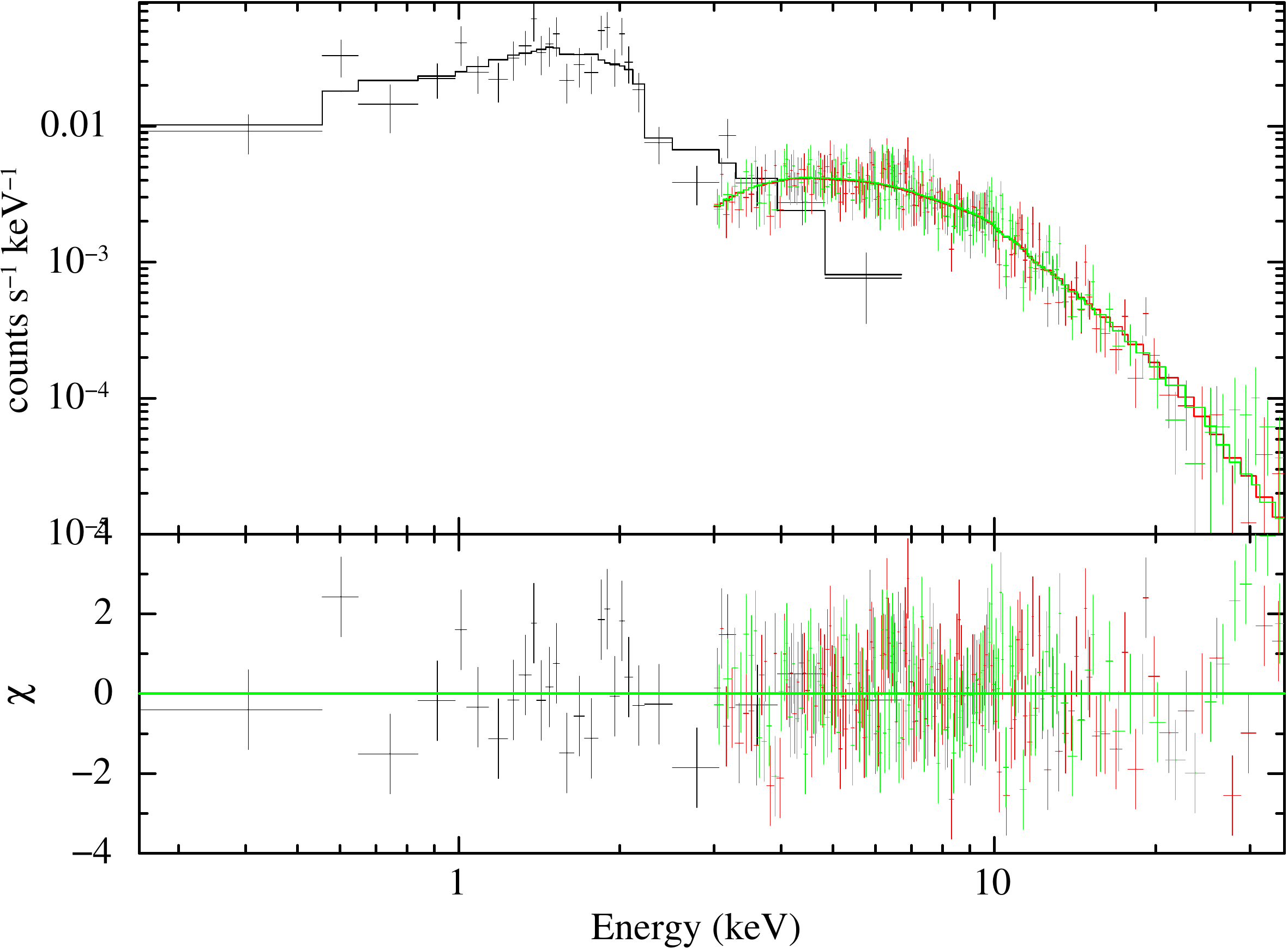}}
 
  \caption{
    Simultaneous fit of \src using the eRASS1 (black) and the \nus data from both the modules; FPMA (red), FPMB (green). The upper panel shows the best-fit spectrum along with the best-fit model (black solid line) and the bottom panel the residuals from the fit. The spectrum has been re-binned only for visual clarity.
  }
  \label{fig:spec-comb}
\end{figure}

The \swift-XRT and the \nus follow-up observations of \src (which were very close in time to epoch 1 of \ero) indicated similar flux levels and no significant variation in the X-ray spectrum.  Motivated by this, we fitted the \ero spectrum from epoch 1 simultaneously with the data from the two \nus modules to constrain the broadband spectral model of \src . 
Constant factors were introduced for each instrument to account for possible absolute cross calibration differences and time variability and were normalised relative to FPMA. 
At first, we tried to apply the best-fit model obtained from the simultaneous fitting of the eRASS spectra (Table~\ref{tcb:eRO_spec}) to the broadband spectrum of the source.
Significant residuals were seen both at the low and high energy ranges indicating that the broadband spectrum cannot be described by the same model. The spectra of HMXBs at first approximation can be described by an absorbed power-law with an exponential roll-over at high energies. Several empirical functional shapes have been used to describe the broadband spectrum of HMXBs (\texttt{cutoffpl}, \texttt{fdcut}, \texttt{NPEX}, etc.). In most of the sources, the rollover energy is found to be in the range of 10--50\,keV. We found an absorbed \texttt{cutoffpl} model to provide a good description of the broadband spectrum of \src. Additional residuals were evident at the low energies and the fit improved significantly by the addition of a partial covering absorber model which is also often required to model the low energy part of the X-ray spectrum of HMXBs \cite[see for e.g.][]{2012MNRAS.420.2307M}. The difference corresponded to $\Delta_{cstat}=15$ for two degrees of freedom. Alternatively, the addition of a blackbody component did not further improve the fit.
The best-fit model is summarised in Table~\ref{tcb:eRO_spec} and the best-fit spectrum is displayed in Fig.~\ref{fig:spec-comb}.

\subsection{The OGLE monitoring of the optical counterpart}
\label{sec:ogle-analysis}

\begin{figure}
\begin{center}
 \resizebox{0.98\hsize}{!}{\includegraphics{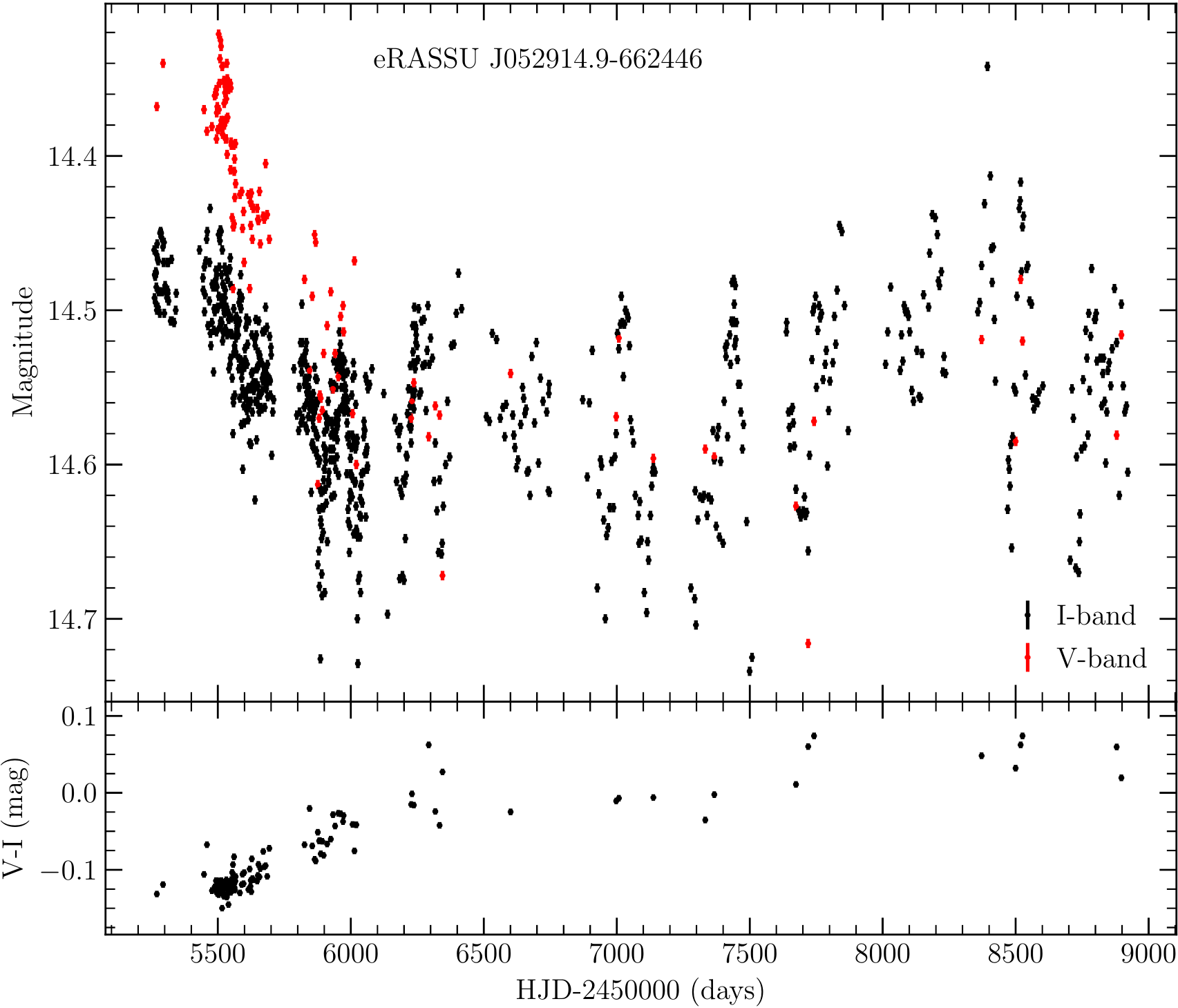}}
\end{center}
  \caption{
    {\it Top:} OGLE-IV light curves in the I- (black) and V-band (red). {\it Bottom:} V-I colour index using I-band data interpolated to the times of the V-band measurements.
  }
  \label{fig:OGLElc}
\end{figure}

\begin{figure}
\begin{center}
 \resizebox{0.9\hsize}{!}{\includegraphics{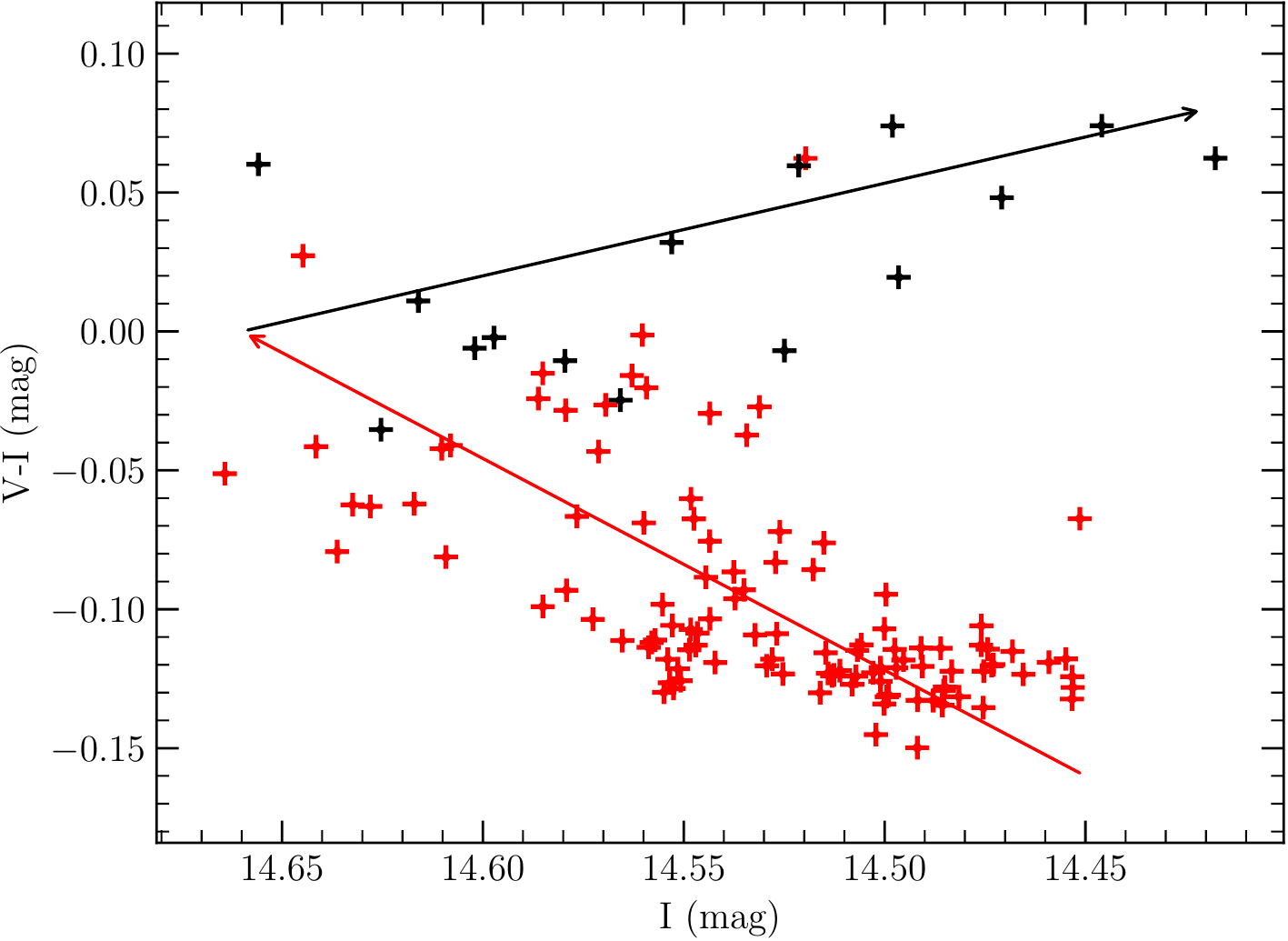}}
\end{center}
  \caption{
    V$-$I colour index as derived for Fig.\,\ref{fig:OGLElc} as function of I-band magnitude. Data taken before HJD 2456500 are marked in red and after in black. The arrows indicate the temporal evolution.
  }
  \label{fig:OGLEVI}
\end{figure}

OGLE observed the optical counterpart of \src for 10 years. The I and V-band light curves are shown in Fig.\,\ref{fig:OGLElc}, together with the temporal evolution of the V$-$I colour index. The light curves show short-term variations on time scales of $\sim$150 days with amplitudes of $\sim$0.2 mag in I, superimposed on a slower brightness change over the full observation period. At the beginning of the monitoring the system was brighter in V than in I, but during the initial brightness decay it faded faster in V, which resulted in a V-I colour index increasing to zero about 1000 days after the beginning of the monitoring. After that the I-band brightness started rising again, but the reddening continued, although at a lower rate  (Fig.\,\ref{fig:OGLElc} bottom panel and Fig.\,\ref{fig:OGLEVI}).

\begin{figure}
\begin{center}

 \resizebox{\hsize}{!}{\includegraphics{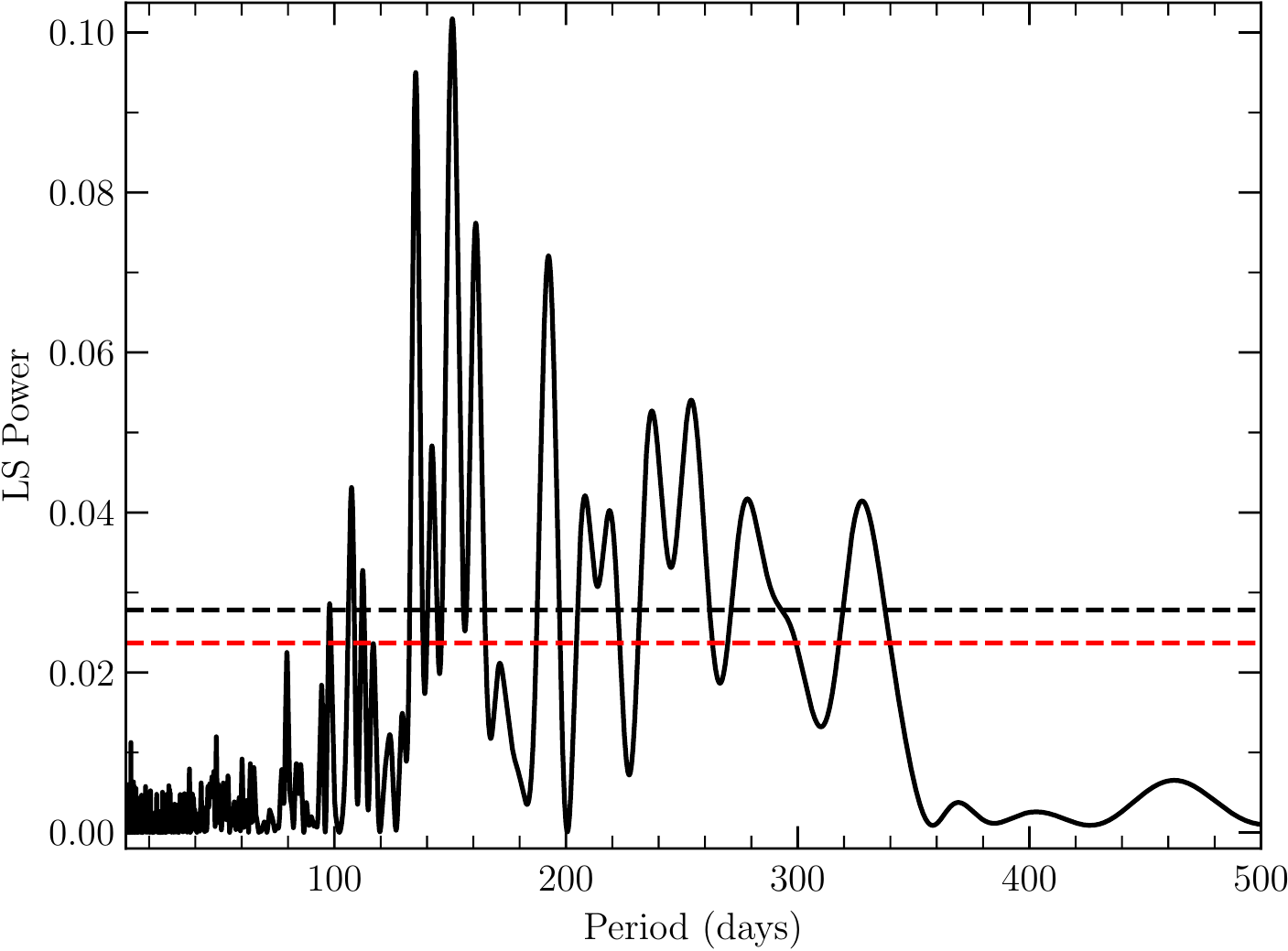}}
 \resizebox{\hsize}{!}{\includegraphics{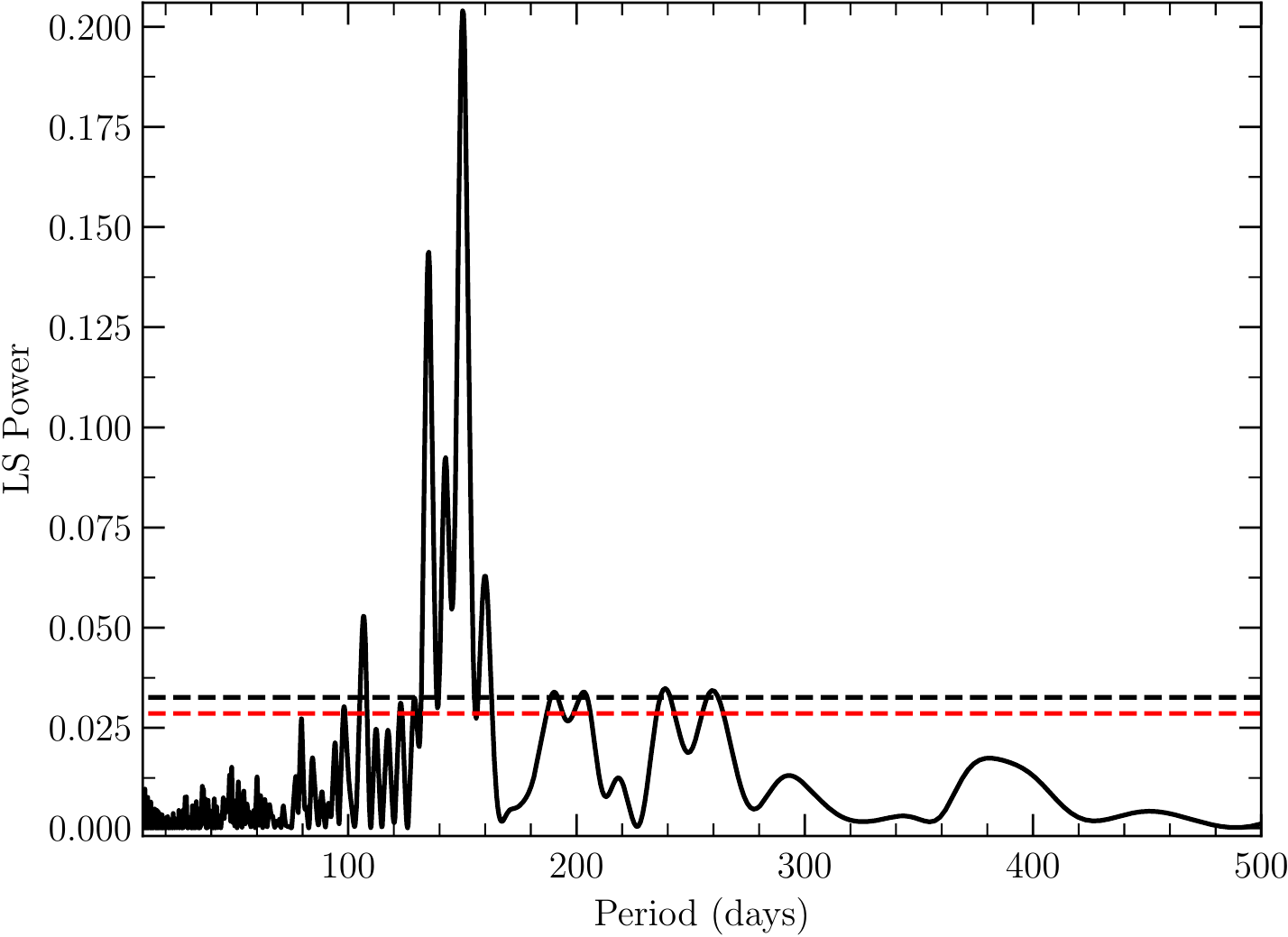}}
\end{center}
  \caption{
    Lomb-Scargle periodograms derived from the measured OGLE I-band light curve (top) and after removing the long-term variation seen in Fig.\,\ref{fig:OGLElc} (bottom). The red and black dashed lines mark the 95\% and 99\% confidence levels. 
  }
  \label{fig:OGLELS}
\end{figure}

\begin{figure}
   \centering
   \resizebox{\hsize}{!}{\includegraphics{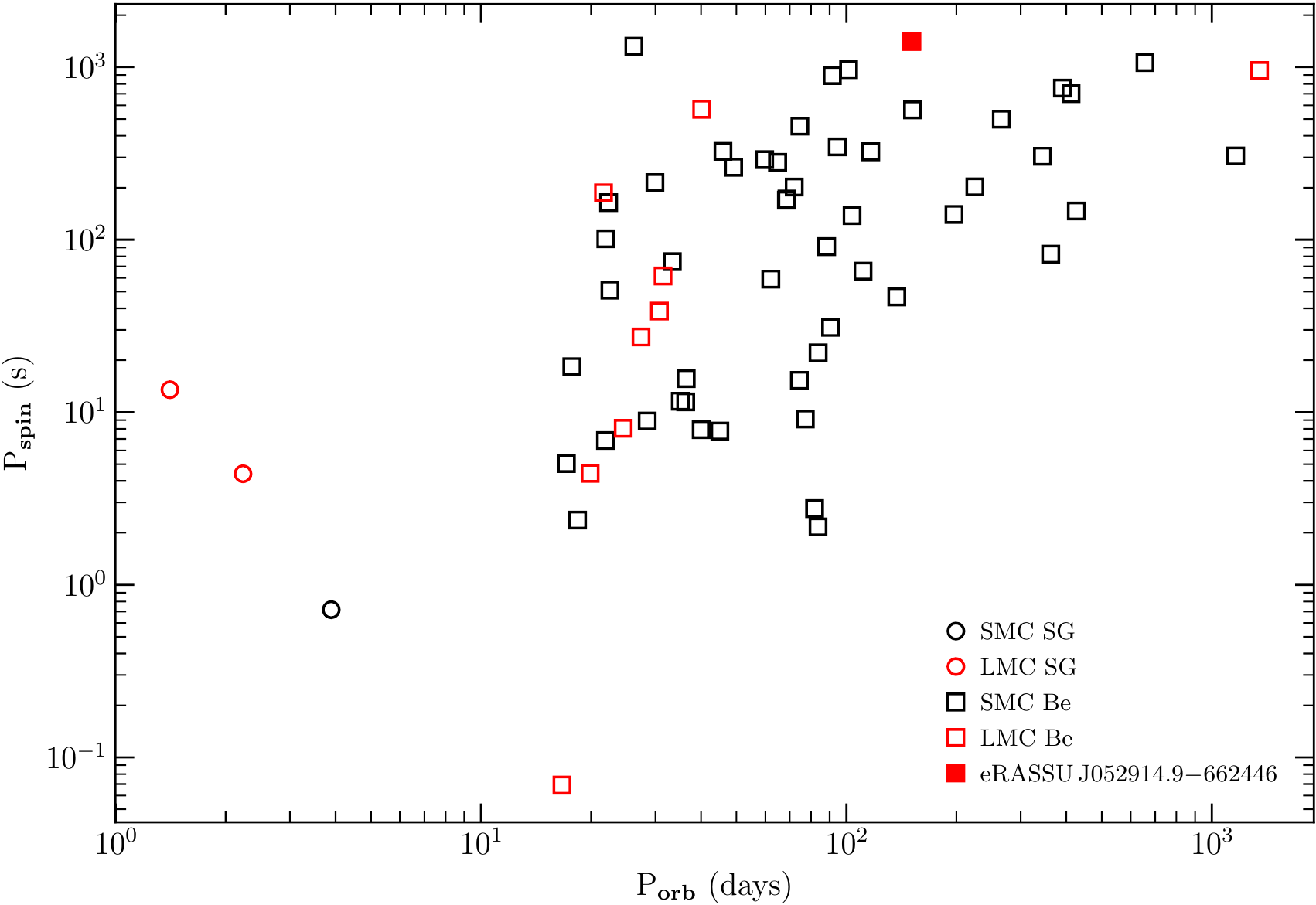}}
   \caption{HMXBs in the Magellanic Clouds: Spin period of the neutron star versus orbital period. BeXRBs and (Roche-lobe filling) supergiant systems are found at different locations in this diagram.} 
   \label{fig:corbet}             
\end{figure}

The OGLE I-band light curve shows short-term variability by about 0.2\,mag on time scales between 100 and 200 days.  
We searched for periodic variations between 20\,days and 500\,days, the typical range of orbital periods found for BeXRBs
\citep{2016A&A...586A..81H}.
The LS periodogram revealed the strongest peak at a period of $\sim$151\,d (Fig.\,\ref{fig:OGLELS}, top), 
but in addition many formally significant peaks due to the various variability time scales in the light curve. Therefore, we subtracted a smoothed light curve 
\citep[derived by applying a Savitzky–Golay filter with a window length of 211 data points;][]{1964AnaCh..36.1627S} to remove the long-term trends \citep[see also][]{2022A&A...662A..22H,2019MNRAS.490.5494M}.
This strongly suppresses the peaks at periods longer than $\sim$180 days in the LS periodogram (Fig.\,\ref{fig:OGLELS}, bottom). However, some peaks between 180 and 300 days remain formally significant, but their broad appearance further indicates an origin from aperiodic variability.
All remaining significant peaks around 151 days can be attributed to aliases with the total length of the OGLE monitoring of 3360 days (peaks at 143 and 160 days) or with seasonal effects (a one year alias at 107 days and an unresolved peak at 135 days caused by three and four years). 
After the detrending of the light curve, the signal of the $\sim$151\,d period strongly increases in significance and 
most likely indicates the orbital period of the binary system (see Fig.~\ref{fig:corbet}).

\subsection{The SALT spectrum}
\label{sec:SALT-analysis}

\begin{figure}
   \centering
   \resizebox{\hsize}{!}{\includegraphics{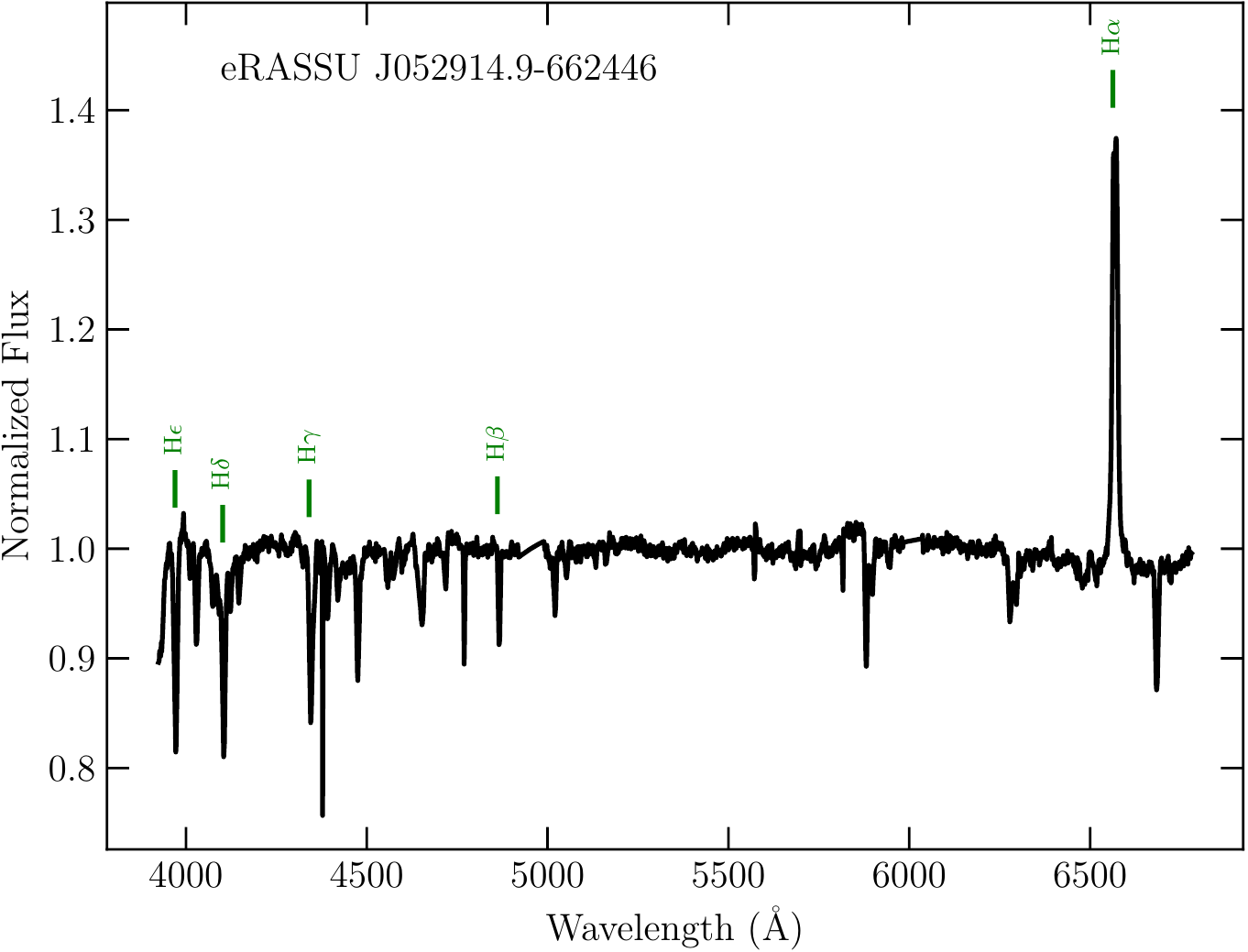}}
   \caption{Optical spectrum of \src obtained on 2020-03-20, using the RSS on SALT. The rest wavelengths of the Balmer lines are marked.}
   \label{fig:SALT}             
\end{figure}

The spectrum obtained by the RSS on SALT is presented in Fig.\,\ref{fig:SALT}.
It is dominated by a double-peaked \Halpha emission line. The measured line equivalent width (EW) is -8.32\,\AA\,$\pm$\,0.37\,\AA\ with a full width at half maximum (FWHM) of 16.72\,\AA\,$\pm$\,0.11\,\AA, which confirms the Be/X-ray binary nature of the source.
From the flux-calibrated SALT spectrum a V-band magnitude of 14.71$\pm$0.05 is inferred.

\section{Discussion}
\label{sec:discussion}
In this paper we report the broadband timing and spectral analysis of the newly discovered source \src through \ero, \swift and \nus data in X-rays and OGLE and SALT RSS data at  optical wavelength. We report in detail the discovery of the spin and the possible orbital period of the system and establish \src as an X-ray binary pulsar. Further, through optical spectroscopic observations and the existence of \Halpha emission the source is identified as a Be X-ray binary pulsar in the LMC. The long spin period of the neutron star places \src at the top of the spin period - orbital period diagram assuming the condition of spin equilibrium \citep[Fig.\,\ref{fig:corbet};][]{1986MNRAS.220.1047C}, for a more recent version see e.g. \citet{2009A&A...505..281M,2019NewAR..8601546K}. The Corbet-diagram also indicates that the LMC BeXRBs seem to lie at the edge of the phase-space occupied by the Magellanic Cloud BeXRBs. The results however might be prone to observational biases due to the relatively smaller number of systems known in the LMC w.r.t. SMC.
The source further adds to the recently growing class of highly magnetised, slowly rotating neutron stars ($\geq$1000\,s), many of which have been discovered in the Magellanic Clouds \citep[][and references therein]{2018MNRAS.475..220V,2021A&A...647A...8M,2020A&A...637A..33T} and helps in understanding the physics of accretion and the interplay between the accretion torque and magnetic field of the NS in these systems as discussed later.

\subsection{Long-term Be-disk/neutron star interaction}
The long-term variability in the X-ray regime covers nearly two decades and indicates that although \src was in the field of view of several \ROSAT PSPC and HRI pointed observations in the 1990's it was detected only twice in $\sim$1995 (MJD 50047) with a flux at least factor of 10 lower than indicated by the recent eRASS surveys.  
The V and I-band variability and the V$-$I colour index also exhibit an interesting pattern over the last decade. The source was detected initially with a much brighter but rapidly decreasing V-band magnitude which decayed by 0.3\,mag on a timescale of $\sim$500 days. Since a major contributor of the V-band brightness is attributed to the inner circumstellar disk \citep{2012ApJ...744...37Y,2012ApJ...753...73Y}, the above occurrence can by explained by the cessation of mass supply from the central Be star causing a depletion in the innermost part of the circumstellar disk and corresponding decay in the V-band brightness. \citet{2001A&A...379..257R,2006A&A...455..953M} and \citet{2016AJ....151..104Y} suggested that the development of a low-density region in the inner circumstellar disk can follow weeks to months after an outburst of the BeXRB.
The above behaviour indicates a possible outburst of \src which might have been missed by X-ray monitoring facilities. Following the rapid decline of the V magnitude, the V$-$I colour index increased to zero and the optical spectrum subsequently 
reddened with a gradually growing contribution of the I-band magnitude. This indicates a growing outer part of the circumstellar disk and an increase in the corresponding disk size. The recent detection of \src in an X-ray bright state by \ero was unfortunately not covered by OGLE monitoring observations. However, the V-band magnitude of 14.71$\pm$0.05, which was obtained from the SALT spectrum is at the faintest level observed during the OGLE monitoring (14.716$\pm$0.003 on HJD 2457719.7, Fig.\,\ref{fig:OGLElc}). The relatively strong \Halpha emission line during this time  suggests that the outer disc was dominant and would have resulted in increased reddening, consistent with the scenario described above.
In addition, the positive correlation between the I-band magnitude and the color suggests that the disk is oriented at low/intermediate inclination angles relative to us \citep[see][and references therein]{2020MNRAS.496.3615M}. Using the relationship between the FWHM and EW of the H-alpha line from Hanuschik et al. (1988), the projected rotational velocity of the star, $v\sin i$, can be estimated: 
\begin{equation}
\log \left[ \frac{FWHM}{2 (v \sin i)} \right] = -0.2\log(-EW) + 0.11
\end{equation}
We obtain $v\sin i \approx 453$ km/s from our measured parameters. \citet[][]{2016A&A...595A.132Z} show that the rotational speeds, $v$ , of Be stars have a range in the fraction of the critical velocity of $0.3 < v/v_c < 0.95$. Assuming an upper limit of the critical velocity of early-type Be stars, $v \approx 600$ km/s \citep[][]{1996MNRAS.280L..31P}, this gives a minimum inclination angle of $i \approx 52^\circ$ consistent with the above scenario.

\subsection{A complex absorber scenario and the non-detection of CRSF for \src}
 The broadband X-ray spectrum is typical of HMXBs and can be described empirically with an absorbed power law with photon index $\sim$1 and an exponential rollover cutoff at $\sim$10\,keV.
The low-energy spectrum (<2\,keV) is however quite complex and can be described by a partial covering absorbing column covering 84\% of the source. The presence of a complex absorber is also supported by the X-ray timing properties of the source, as the morphology of the pulse profile (Fig.\,\ref{fig:pp}) exhibits a narrow (0.1 in phase) dip at phase $\sim$0.7.
Pulse profiles of Be/X-ray binary pulsars are known to be complex in low ($<$10\,keV) X-ray energy ranges with
multiple emission peaks and sometimes narrow absorption dips. Phase-resolved spectroscopy in some of the Galactic BeXRBs revealed that matter in the accretion streams partially obscures the emitted radiation causing the dips \citep[e.g.][]{2012MNRAS.420.2307M,2013ApJ...764..158N}.
Examples in the Magellanic Clouds are 
Swift\,J053041.9$-$665426 = LXP\,28.8 \citep{2013A&A...558A..74V}, XMMU\,J005929.0$-$723703 = SXP\,202b \citep[][]{2008A&A...489..327H} and XMMU\,J045736.9$-$692727\xspace \citep[][]{2022arXiv220300625H}. 

 The magnetic field strength of the neutron star can be robustly determined from an observed cyclotron resonance scattering feature (CRSF) centroid energy as
$E_{\mathrm{cyc}}$ (determined from Obs. 1), and is given as:
\begin{equation}
	E_{\mathrm{cyc}} = \frac{11.57\,\mathrm{keV}}{1+z} \times B_{12}
	\label{eqn:12b12}
\end{equation}
where $B_{12}$ is the field strength in units of $10^{12}$\,G; $z \sim 0.3$ is the gravitational redshift in the scattering region for standard neutron star parameters.
CRSFs are often observed as broad, shallow features that can be difficult to detect against the continuum spectrum. Although we did not find any evidence for a statistically significant feature in the X-ray spectrum, we computed an upper limit on the depth of a putative CRSF by varying the line centroid energy in the range of 7-25\,keV (energy range where the signal to noise is the highest) in steps of 500\,eV and by fixing the width at 20\% its centroid energy \citep[see for e.g.][for typically observed energy to width ratios]{2017JApA...38...50M,2019A&A...622A..61S}. We obtained a highest line depth of $\tau_{\rm c} \sim0.8$ at an energy of 17\,keV. The CRSF energy would imply a magnetic field strength of  $B\approx10^{12}$\ G for the NS.

\subsection{Subsonic accretion onto a slowly rotating highly magnetised neutron star in \src}

A plausible explanation for low X-ray luminosity ($\sim$\oexpo{35}--\oergs{36}) and slowly rotating ($\sim$1000\,s) neutron stars is that these systems may be quasi-spherically accreting from stellar winds. In this regime, subsonic or settling accretion occurs when the plasma remains hot until it meets the magnetospheric boundary. A hot quasi-spherical shell is formed around the magnetosphere  as the Compton cooling timescale of plasma above the magnetosphere is much smaller then the timescale of the freely falling fresh matter that is gravitationally captured from the stellar wind \citep{2012MNRAS.420..216S}.  The actual accretion rate onto the neutron star is determined by the ability of the plasma to enter the magnetosphere due to the Rayleigh-Taylor instability and varies between 0.1--0.5 of the Bondi mass accretion rate \citep[][]{2012MNRAS.420..216S,2017arXiv170203393S,2018ASSL..454..331S}. The shell mediates the angular momentum transport to or from the rotating neutron star and both spin-up and spin-down is possible depending on the sign of the angular momentum difference between the matter being accreted and the magnetospheric boundary. Under the assumption of spin equilibrium the spin period can be derived as in \citet{2017arXiv170203393S}  ignoring  the dimensionless theory parameters $\Pi_0$, $\zeta$ of about unity, we find
\begin{equation}
\label{e:Peq}
    P_\mathrm{eq}\approx 1000[\mathrm{s}]~\mu_{30}^{12/11}
    \left(\frac{P_\mathrm{b}}{10\mathrm{\,d}}\right)\dot M_{16}^{-4/11}v_8^4\,.
\end{equation}
Here the parameters are normalised to the typical values; $\mu_{30} \equiv \mu$ / (10$^{30}$\,G\,cm$^3$), 
$P_\mathrm{b}$  is the binary orbital period (the orbit is assumed to be circular), 
$\dot M_{16}\equiv \dot M$ / (10$^{16}$\,g\,s$^{-1}$) is the accretion rate onto the neutron star surface corresponding to an X-ray luminosity of $L_x=0.15\dot Mc^2$, and 
$v_8\equiv v$ / (1000\,km\,s$^{-1}$) is the stellar wind velocity relative to the neutron star. 
Substituting $P\approx 1412$\,s, $\dot M_{16}\simeq 7$, $P_\mathrm{b}=151$\,d we find
\begin{equation}
\label{e:mu1323}
    \mu_{30}\simeq 0.2 
     v_8^{-11/3}\,.
\end{equation}
We note that the estimate of the  magnetic field of the neutron star is strongly dependent on the stellar wind velocity. For a value of 1000\,km\,s$^{-1}$, 
$B\approx 10^{12}$~G indicating a typical magnetic field strength of a NS.

It should be noted that the main source of uncertainty in the estimation of $B$ using the quasi-spherical accreting scenario is from the wind velocity.
The stellar wind of a Be star is highly anisotropic in nature. At its poles wind velocities are high but the density is low, while it is denser near the equator but much slower due to the presence of the disk \citep{1978ApJS...38..229P,1991A&A...244L...5L}. \citet{1989A&A...223..196W} provide a prescription of estimating the typical wind velocity of BeXRBs as described below. From the measured 
 \Halpha equivalent width (Sect.~\ref{sec:SALT-analysis}) of -8.32\,\AA\,$\pm$\,0.37\,\AA\ one can estimate the size of the Be disk using the relation from \citet{2006ApJ...651L..53G,2007ApJ...656..437G} to be $R\sim4\,R_\mathrm{s}$, where $R_\mathrm{s}$ is the radius of the star. 
 Further, following equations 4 and 5 of \citet{1989A&A...223..196W} one can derive an upper limit on the radial velocity $v_\mathrm{r}$ component and the rotational wind  $v_{w}$ component as $120\,\mathrm{km}\,\mathrm{s}^{-1}$ and $74\,\mathrm{km}\,\mathrm{s}^{-1}$, 
respectively, assuming maximum values for $v_{0}=30\,\mathrm{km}\,\mathrm{s}^{-1}$, $n=3.2$, and $v_{eq}=300\,\mathrm{km}\,\mathrm{s}^{-1}$ \citep[see][for a range of rotational velocity of Be stars observed in the LMC]{2004PASA...21..310K}. This leads to a value of $v_{rw}\sim 150\,\mathrm{km}\,\mathrm{s}^{-1}$ for the net wind velocity component.
For computing the wind velocity relative to the NS one needs to take into account the orbital motion. Assuming P$_{\mathrm b}$\,=\,151 days and a Be star with mass $\approx$10\,\msun yields a velocity also of order $\approx$100\,\kms assuming a circular orbit with zero eccentricity. This leads to an upper limit of $\approx$250\,\kms relative velocity. However a majority of the observed BeXRBs have eccentric orbits. In order to account for a possible eccentricity of the orbit in \src, we looked at the sample of BeXRBs in the SMC with reliably measured eccentricity values \citep[][]{2011MNRAS.416.1556T}. Assuming a highly eccentric orbit from the maximum measured value of $\mathrm{e}=0.8$ leads to a maximum velocity of $\approx$300\,\kms at perigee and an upper limit of the relative velocity of $\approx$450\,\kms. The above values lead to a lower limit of the magnetic field strength between $4\times10^{13}-10^{14}$~G for the subsonic accretion scenario indicating a highly magnetised NS or a magnetar.


It is also worth mentioning that the magnetic field strength assuming disk accretion and using the standard prescription of \citet{1979ApJ...234..296G} under the equilibrium hypothesis leads to an estimation of a factor of  $\sim$10 or higher $B$ for such slowly rotating pulsars. Further, future monitoring programs are essential for determining the validity of the assumption of spin equilibrium of the system, and whether the pulsar is spinning up or down.
\section{Conclusions}
\label{sec:conclusion}
The main results can be summarised as follows:
\begin{itemize}
    \item We report the discovery of pulsations at a period of 1412\,s  from the newly discovered source \src in the LMC. Pulsations are detected in a broad energy range of 0.2--20\,keV and the pulse profile does not show strong energy dependence.
    \item The X-ray position is consistent with a star with magnitudes and colours indicative for an early spectral type.  Further, our optical spectroscopic observations revealed the existence of \Halpha emission and the source is identified as a Be X-ray binary pulsar in the LMC. 
    
    \item The broadband X-ray spectrum is typical of HMXBs and can be described empirically with an absorbed power law with photon index $\sim$1 and an exponential rollover cutoff at $\sim$10\,keV. No evidence of a cyclotron resonance scattering feature (CRSF) was found in the broadband X-ray spectrum. We obtained an upper limit of $\tau_{\rm c} \sim0.8$ on the depth of a CRSF at a centroid energy of 17\,keV for a typical magnetic field strength of $\approx10^{12}$\,G for a NS. The estimation of B under assumption of spin equilibrium for a quasi-spherical subsonic accretion scenario for \src indicates a much higher B value of $4\times10^{13}-10^{14}$~G for the NS but is prone to several uncertainties, especially in the estimation of the wind velocity.
    
    \item Decadal monitoring with OGLE revealed a variable optical counterpart with a periodicity of $\sim$151\,days in the I-band, which indicates the orbital period of the system. At the beginning of the optical monitoring the system was brighter in V which decayed very rapidly, resulting in a V$-$I colour index increasing to zero, and after the I-band brightness started rising again the reddening of the system continued. \src was serendipitously located in the field of view of several \ROSAT PSPC and HRI pointed observations and was detected only twice in $\sim$1995 (MJD 50047) with a flux at least factor of 10 lower than indicated by the recent eRASS surveys suggesting a variable nature in the X-ray regime. 
\end{itemize}

\bibliographystyle{aa} 
\bibliography{references} 

\begin{acknowledgements}
We thank the referee for useful comments and suggestions. 
This work is based on data from \ero, the soft X-ray instrument aboard \srg, a joint Russian-German science mission supported by the Russian Space Agency (Roskosmos), in the interests of the Russian Academy of Sciences represented by its Space Research Institute (IKI), and the Deutsches Zentrum f{\"u}r Luft- und Raumfahrt (DLR). The \srg spacecraft was built by Lavochkin Association (NPOL) and its subcontractors, and is operated by NPOL with support from the Max Planck Institute for Extraterrestrial Physics (MPE).
The development and construction of the \ero X-ray instrument was led by MPE, with contributions from the Dr.~Karl Remeis-Observatory Bamberg \& ECAP (FAU Erlangen-N{\"u}rnberg), the University of Hamburg Observatory, the Leibniz Institute for Astrophysics Potsdam (AIP), and the Institute for Astronomy and Astrophysics of the University of T{\"u}bingen, with the support of DLR and the Max Planck Society. The Argelander Institute for Astronomy of the University of Bonn and the Ludwig Maximilians Universit{\"a}t Munich also participated in the science preparation for \ero.
The \ero data shown here were processed using the \eSASS software system developed by the German \ero consortium.

This work used observations obtained with \xmm, an ESA science mission with instruments and contributions directly funded by ESA Member States and NASA. The \xmm project is supported by the DLR and the Max Planck Society. This work is partially supported by the \textsl{Bundesministerium f\"{u}r Wirtschaft und Energie} through the \textsl{Deutsches Zentrum f\"{u}r Luft- und Raumfahrt e.V. (DLR)} under the grant number FKZ 50 QR 2102. 

This research has made use of the VizieR catalogue access tool, CDS,
Strasbourg, France. The original description of the VizieR service was
published in A\&AS 143, 23.
\end{acknowledgements}

\end{document}